\documentclass[12pt,preprint]{aastex}
\usepackage{epsfig}
\usepackage{lscape}
\usepackage{rotating}


\def\ls{{_<\atop^{\sim}}}
\def\gs{{_>\atop^{\sim}}}

\lefthead{Nicastro et al.}

\begin{document}

\title{Chandra Detection of the First X-ray Forest along the Line of 
Sight To Mkn~421}

\author{Fabrizio Nicastro,$\!$\altaffilmark{1}, 
Smita Mathur$\!$\altaffilmark{2}, 
Martin Elvis$\!$\altaffilmark{1}, 
Jeremy Drake$\!$\altaffilmark{1}
Fabrizio Fiore$\!$\altaffilmark{3},
Taotao Fang$\!$\altaffilmark{4},
Antonella Fruscione$\!$\altaffilmark{1},
Yair Krongold$\!$\altaffilmark{1}
Herman Marshall$\!$\altaffilmark{5},
Rik Williams$\!$\altaffilmark{2}}

\altaffiltext{1}{Harvard-Smithsonian Center for Astrophysics,
60 Garden st., Cambridge, MA 02138, USA}
\altaffiltext{2}{Ohio-State University, Columbus, OH, USA}
\altaffiltext{3}{Osservatorio Astronomico di Roma, Monteporzio, RM, Italy}
\altaffiltext{4}{University of California-Berkeley, Berkeley, CA, USA}
\altaffiltext{5}{Massachusetts Institute of Technology, Cambridge, MA, USA}


\begin{abstract}
We present the first $\ge 3.5\sigma$ (conservative) or $\ge 5.8\sigma$ 
(sum of lines significance) detection of two Warm-Hot Intergalactic Medium 
(WHIM) filaments at $z>0$, which we find along the line of sight to the 
blazar Mkn~421. 
These systems are detected through highly ionized resonant metal absorption 
in high quality {\em Chandra}-ACIS and -HRC Low Energy Transmission Grating 
(LETG) spectra of Mkn~421, obtained following our two Target of Opportunity 
requests during two outburst phases ($F_{0.5-2 keV} = 40$ and 60 mCrab; 1 
Crab = 2$\times 10^{-8}$ erg s$^{-1}$ cm$^{-2}$). 

Columns of He-like oxygen and H-like nitrogen can be detected in the 
coadded LETG spectrum of Mkn~421 down to a sensitivity of N$_{OVII} \ge 8 
\times 10^{14}$ cm$^{-2}$ and N$_{NVII} \ge 10^{15}$ cm$^{-2}$ respectively, 
at a significance $\ge 3\sigma$. The two intervening WHIM systems that we 
detect, have OVII and NVII columns of N$_{OVII} = (1.0 \pm 0.3) \times 
10^{15}$ cm$^{-2}$ N$_{NVII} = (0.8 \pm 0.4) \times 10^{15}$ cm$^{-2}$, and 
N$_{OVII} = (0.7 \pm 0.3) \times 10^{15}$ cm$^{-2}$, N$_{NVII} = 
(1.4 \pm 0.5) \times 10^{15}$ cm$^{-2}$ respectively. 
We identify the closest of these two systems with an intervening WHIM 
filament at $cz = 3300 \pm 300$ km s$^{-1}$. The second system, instead, 
at $cz = 8090 \pm 300$ km s$^{-1}$, is identified with an intervening WHIM 
filament located $\sim 13$ Mpc from the Blazar. 
The filament at $cz = 3300 \pm 300$ lies $\ls 5$ Mpc from a known HI 
Ly$\alpha$ system at $cz = (3046 \pm 12)$ km s$^{-1}$ (Shull et al., 1996) 
whose 3$\sigma$ maximal HI kinetic temperature, as derived from the 
observed line FWHM, 
is $T \le 1.2 \times 10^5$ K. This temperature is inconsistent with the 
temperature measured for the X-ray filament, so if the systems are related 
a multiphase WHIM is required. 

Combining UV and FUV upper limits on the HI Ly$\alpha$ and the 
OVI$_{2s\rightarrow 2p}$ transitions, with our measurements in the 
X-rays, we show that, for both filaments, equilibrium collisional ionization 
(with residual photoionization by both the diffuse UV and X-ray background, 
and the beamed emission of Mkn~421 along our line of sight) provides 
acceptable solutions. These solutions define ranges of temperatures, 
metallicity ratios and equivalent H column densities, which are in good 
agreement with the predictions of hydrodynamical simulations for the 
formation of large scale structures in the Universe. 

From the detected number of WHIM filaments along this line of sight we 
can estimate the number of OVII filaments per unit redshift with columns 
larger than $7 \times 10^{14}$ cm$^{-2}$, $d\mathcal{N}_{OVII}/dz(N_{OVII} 
\ge 7 \times 10^{14}) = 67^{+88}_{-43}$, consistent, within 
the large 1$\sigma$ errors, with the hydrodynamical simulation predictions 
of $d\mathcal{N}_{OVII}/dz(N_{OVII} \ge 7 \times 10^{14}) = 30$. 
Finally, we measure a cosmological mass density of X-ray WHIM filaments 
$\Omega_b = 0.027^{+0.038}_{-0.019} \times 10^{[O/H]_{-1}}$, consistent with 
both model predictions and the estimated number of 'missing' baryons at 
low redshift. 
\end{abstract}


\section{Introduction}
Despite dramatic recent progress in cosmology, very little is yet known 
about the location of the normal baryonic matter in the local Universe. 
The extraordinary WMAP ({\em Wilkinson Microwave Anisotropy Probe}) 
measurements of the Cosmic Microwave Background (CMB) anisotropies favor a 
$\Lambda$-CDM scenario in which the baryon density in the Universe 
amounts to $\Omega_b = (4.6 \pm 0.2) h_{70}^{-2}$ \% of the total 
matter-energy density, with a baryon-to-matter ratio of $0.17 \pm 0.01$ 
(Bennett et al., 2003, Spergel et al., 2003). 
This number agrees well with predictions 
by the standard 'big bang nucleosynthesis' when combined with light element 
ratios ($\Omega_b = (4.4 \pm 0.4) h_{70}^{-2}$ \%: Kirkman et al., 2003), 
and also with the actual number of baryons detected at redshifts 
larger than 2 in the 'trees' of the Ly$\alpha$ Forest: 
$\Omega_b \ge 3.5 h_{70}^{-2}$ \% (Rauch, 1998; Weinberg et al., 
1997). This concordance cosmology represents a major advance. However, 
the number of baryons actually detected in the virialized Universe 
(i.e. stars, neutral H and He and molecular H associated with galaxies, 
X-ray emitting gas in clusters and big groups with M$\ge 4 \times 10^{13}$ 
M$_{\odot}$: Fukugita, 2003) and the non yet virialized Universe (i.e. 
photoionized warm/cool gas in the local Lyman-$\alpha$ forest filaments 
and shock-heated warm-hot gas in the OVI filaments: Shull, 2003; Stocke, 
Shull \& Penton, 2004; Danforth \& Shull, 2005), is less than half of the 
expected concordance value, 
$\Omega_b(detected) = (2.1 \pm 0.3)$ \% (for $h_{70} = 1$). 
The number of 'missing baryons' at $z \ls 2$ is therefore $\Omega_b(missing) 
= (2.5 \pm 0.4)$ \%. 

If we cannot find half of the ordinary matter in our local Universe it does 
not bode well for understanding dark matter or dark energy. 

\medskip
Hydrodynamical simulations for the formation of structures in the 
Universe, offer a possible solution to this puzzle of the ``missing 
baryons''. 
They predict that in the present epoch ($z \ls 1-2$), 30-40 \% of the 
ordinary, baryonic, matter in the Universe lies in a tenuous medium at 
overdensities $\delta \simeq 5-200$
\footnote{$\delta = (n_b / <n_b>)$ is the 
baryon gas overdensity in units of average density in the Universe, 
$<n_b> = 2 \times 10^{-7} (1 + z)^3 (\Omega_b h^2/0.02)$ cm$^{-3}$.}
, relative to the mean baryon density 
in the Universe ($\sim 10^6$ below the density of the interstellar medium 
in our Galaxy). This medium is in a ``warm-hot'' phase and forms the 
so-called 'Warm Hot Intergalactic Medium' or ``WHIM''. The WHIM was 
shock-heated to temperatures of $10^5-10^7$ K during the continuous 
process of structure formation (e.g. Hellsten et al., 1998; Cen \& 
Ostriker, 1999; Dav\'e et al., 2001; Fang, Bryan \& Canizares, , 2002). 
At these temperatures C, N, O and Ne (the most abundant metals in gas with 
solar or solar-like composition), are mainly distributed between their 
He-like and H-like species. The strongest bound-bound transitions from 
these ions fall in the soft X-ray band (e.g $\lambda(OVII_{K\alpha}) = 
21.602$ \AA, and $\lambda(OVIII_{K\alpha} = 18.97$ \AA), and so a 'forest' 
of weak (i.e. N$_{X^i} < 10^{16}$ cm$^{-2}$, or $W_{X^i}$ $< 18-30$ m\AA\, 
depending on the particular ion
\footnote{here, and throughout the paper, $X^i$ is the generic He- or H-like 
ion of C, N, O or Ne}
) metal absorption lines is expected to be imprinted in the X-ray spectra of 
all background sources at redshifts between zero and $\sim 1-2$ , by the WHIM 
filaments (e.g. Perna \& Loeb, 1998; Hellsten et al., 1998; Fang, Bryan \& 
Canizares, 2002). 
These lines are the direct analogous of the 'Lyman-$\alpha$ forest' due to 
cool HI and detected copiously at higher redshift in the optical spectra of 
quasars. 

Ubiquitous high-velocity OVI absorption (Sembach et al., 2003) and 
higher-ionization, CV-VI, NVI-VII, OVII-VIII and NeIX-X absorption 
(Nicastro et al., 2002, Fang et al., 2003, Rasmussen et al., 2003, 
Cagnoni et al., 2004, Williams et al., 2005) at $z\simeq0$ have been 
detected in the far-ultraviolet (FUV) and X-ray spectra of extragalactic 
sources. The exact location of this ubiquitous absorber, whether in an 
extended Galactic corona (Sembach et al., 2003, Collins et al., 2004) or in 
a local multiphase WHIM filament containing our own Local Group of Galaxies 
(Nicastro et al., 2002, Nicastro et al., 2003, Williams et al., 2005) is 
still highly controversial. Either case the success in 
detecting such a diffuse medium at $z=0$ is due to our special embedded 
location and so on the conspicuous portion of absorber crossed by any single 
extragalactic line of sight. 

Detecting the WHIM outside the Local Group, instead, means observing 
along a random, unprivileged line of sight. 
Due to the intrinsic steepness of the 'Log$\cal{N}$-Log(N$_X^i$)' 
of the WHIM metal absorption lines, at relatively large columns 
(e.g. Hellsten et al., 1998, Fang, Bryan \& Canizares, 2002), only a single 
OVII K$\alpha$ system with N$_{OVII} \ge 10^{16}$ cm$^{-2}$ is expected along 
a random line of sight, up to $z = 0.3$, with a probability of $\sim 60$ \%. 
This number rises to 8 for systems with N$_{OVII} \times 10^{15}$ cm$^{-2}$. 
So detecting weaker lines is expected to be more productive than probing 
higher redshifts. 

Consistently with expectations, no secure X-ray detection of $z > 0$ 
intervening WHIM absorption systems has been claimed so far. 
The deepest observation of a quasar at $z\sim 0.3$ is the 
500 ks {\em Chandra}-LETG observation of H~1821+643. 
This observation is sensitive only to N$_{X^i} \gs 7 \times 
10^{15}$ cm$^{-2}$ (at $\ge 3 \sigma$ level), and only 2-3 low-significance 
($\sim 1-2 \sigma$) detections of such systems have been claimed 
(Mathur, Weinberg, \& Chen, 2003). 
Ravasio et al. (2005) recently published a moderate signal to noise 
{\em Newton}-XMM Reflection Grating Spectrometer (RGS) observation of 
the neraby ($z=0.03$) of Mkn~421. This spectrum is sensitive to OVII 
columns of $\sim 3-4 \times 10^{15}$ cm$^{-2}$, at 3$\sigma$ and no 
intervening systems could be identified down to this OVII column 
sensity threshold. 
Finally, another $z > 0$ OVIII WHIM candidate, at $z = 0.0554$ was claimed by 
Fang et al. (2002), in a {\em Chandra} Advanced CCD Imaging Spectrometer 
(ACIS) LETG spectrum of the blazar PKS~2155-304. 
However neither a deeper {\em Chandra} High Resolution Camera (HRC) LETG 
nor a {\em Newton}-XMM observation of the same object could confirm 
this result (Nicastro et al., 2002, Rassmussen et al., 2003, 
Cagnoni et al., 2004). 
Clearly higher signal-to-noise spectra are urgently needed. 

A very large number of counts per resolution element (CPREs) is needed 
in current {\em Chandra} Low Energy Transmission Grating (LETG) or 
{\em Newton}-XMM Reflection Grating Spectrometer (RGS) spectra, both 
of which have a resolution of R$\sim 400$ at $\sim 20$ \AA, to detect these 
lines. 
In an ideal detector about 30 counts, at 21.6 \AA, are 
needed in a $\Delta \lambda = 50$ m\AA\ continuum bin (i.e. 1 resolution 
element) for a 3$\sigma$ detection of a 29 m\AA\ absorption line. 
In practice, at such a low number of CPREs 
and for moderately noisy detectors (like the {\em Chandra} HRCS-LETG), at 
least 100 CPREs are needed.  Such a line probes OVII columns of about 
$10^{16}$ cm$^{-2}$, the very high-column tail of the WHIM distribution. 
To probe the more common WHIM columns of $N_{OVII} \sim 10^{15}$ cm$^{-2}$, 
about 2500 CPREs are needed for a 3$\sigma$ 
detection. 
The brightest Seyfert galaxies and blazars have typical 
soft X-ray (0.5-2 keV) fluxes of the order of $\sim 4 \times 10^{-11}$ erg 
s$^{-1}$ cm$^{-2}$ (2 mCrab), which would take a 2.5 Ms {\em Chandra}-LETG 
exposure to reach this S/N. 
Furthermore all these objects are nearby ($z\ls 0.05$) and so sample only  
path-lengths of a few hundred Mpc. 

\medskip
A possible solution is to observe background sources in untypically bright 
states. This strategy has been previously and successfully exploited 
with low resolution spectrometers (e.g. {\em Beppo}SAX) to study 
Gamma Ray Burst X-ray afterglow, but also compact Galactic sources 
and non-beamed AGNs (e.g. Puccetti et al., 2005, in preparation).
We have extended this strategy to beamed, pure continuum, AGNs in order 
to obtain exceptionally good quality spectra of the intervening 
Intergalactic Medium (IGM). 
In this paper we present the first results of this highly successful 
observational strategy. 
We report the first high significance ($\sigma = 3.5$ and $\sigma = 4.8$) 
detections of two high ionization absorbers at cosmological distances, which 
we identify with WHIM filaments, in a high signal-to-noise {\em Chandra}-LETG 
spectrum of the blazar Mkn~421. 
A companion, complementary paper, which reports on the first estimate of 
the cosmological mass density of OVII WHIM filaments derived from these 
detections, has been recently published (Nicastro et al., 2005). 

Throughout the paper we use LETG spectra grouped at resolutions of 
$\Delta \lambda = 12.5$ m\AA\ (for fitting purposes) and 25 m\AA\ (for 
plotting purposes), i.e. respectively 4 and 2 times better than the 
intrinsic LETG resolution of 50 m\AA. 
We adopt $H_0 = 72$ km s$^{-1}$ Mpc$^{-1}$ (Freedman et al., 2001, 
Bennett et al., 2003). 
Errors are quoted at 1$\sigma$ for 1 interesting parameters, unless 
otherwise stated. 

\section{The {\em Chandra} Data of Mkn~421}
Mkn~421 is a blazar at $z=0.03$ (De Vaucouleurs et al., 1991), which undergoes 
frequent strong 
outbursts from its quiescent state of $\sim 1$ mCrab ($2 \times 10^{-11}$ 
erg s$^{-1}$ cm$^{-2}$) to 20-50 mCrab (Figure 1). 
When in outburst 
Mkn~421 is one of the few objects detected at TeV energies 
(e.g. Okumura et al., 2002). 
Mkn~421 has been observed by {\em Chandra} at four different epochs (Figure 1, 
Table 1). 
%
\begin{table}
\footnotesize
\begin{center}
\caption{\bf \small Table of X-ray Observations} 
\vspace{0.4truecm}
\begin{tabular}{|ccccccc|}
\hline
Date & MJD & Det/Grating & Exposure & (0.5-2) keV & Total Dispersed & CPREs \\
& & & & Flux & Counts & at 21 \AA\ \\
\hline
& & & ks & mCrab & & \\
\hline
11/05/99 & 51487 & ACIS/HETG & 25.8 & 0.5 & 9635$^a$ & 2$^a$ \\
05/29/00 & 51693 & ACIS/HETG & 19.6 & 9.5 & 125301$^a$ & 17$^a$ \\
05/29/00 & 51693 & HRC/LETG & 19.7 & 13.5 & 209586$^b$ & 177$^b$ \\
10/26/02 & 52573 & ACIS/LETG & 91.5 & 60.0 & 4213181$^c$ & 2676$^c$ \\ 
07/01/03 & 52821 & HRC/LETG & 99.2 & 40.0 & 2967358$^b$ & 2512$^b$ \\
\hline 
\end{tabular}
\end{center}
$^a$ 1st order MEG. 
$^b$ All orders LETG. 
$^c$ 1st order LETG. 
\end{table}
\normalsize
%
The first short observation was performed with the ACIS 
High Energy Transmission Grating (HETG) configuration, at the beginning of 
the satellite calibration phase, on 1999 November 5. 
This observation lasted 25.8 ks, and the target was caught at 
a historical minimum in its 0.5-2 keV flux, of $F_{0.5=2 keV} = 10^{-11}$ 
erg s$^{-1}$ cm$^{-2} = 0.5$ mCrab, or 0.75 mCrab extrapolating to the 
1.5-12 keV band of the {\em Rossi}-XTE All Sky Monitor (ASM: e.g. 
Levine et al., 1996: Figure 1). 
The number of photons per resolution elements ($\Delta \lambda = 20$ m\AA) 
in the 1st-order HETG spectrum of Mkn~421 was therefore far below the 
threshold needed for detection of WHIM absorption lines, and so useless 
for our purposes (Table 1). We no longer consider this observation here.  
Six months later (2000, May 29) we requested and obtained two short 
consecutive {\em Chandra} Director Discretionary Time (DDT) observations 
of Mkn~421, following 2 weeks of intense activity of the target as recorded 
by the ASM (Figure 1). 
The two observations were performed $\sim 1$ week after the trigger with the 
{\em Chandra} ACIS-HETG and HRC-LETG, and lasted $\sim 20$ ks each, recording 
0.5-2 keV fluxes some 20 times higher than before, 9.5 and 13.5 mCrab 
respectively (Figure 1, Table 1). 
Despite the large fluxes, only 177 and 17 CPREs were detected at $\lambda = 21.0$ \AA\ in the LETG (all-orders) 
and HETG (1st-order) spectra, respectively (Table 1). In the following we will 
use only the HRC-LETG spectrum and only in combination with the subsequent, 
much better quality, LETG spectra of the source. 
Finally, two further 100 ks {\em Chandra} ACIS-LETG and 
HRC-LETG observations of Mkn~421 were performed in 2002, October, 
26-27 (Figure 1 and 2, Table 1), and 2003, July 1-2 (Figure 1, Table 1), under our 
approved {\em Chandra}-AO4 program to observe blazars in outburst phases. 
These two observations were performed $\sim 24$ and $\sim 18$ hours respectively 
after the trigger and caught the source at a historical maximum, 
$\sim 100$ times brighter than in the first observation, 60 and 40 mCrab 
in the 0.5-2 keV band (Figure 1, Table 1), and allowed us to collect 
$\sim 2600$ CPREs each at 21 \AA\ (Table 1). 
%
\begin{figure}
\hspace{-0.8in}
\epsfbox{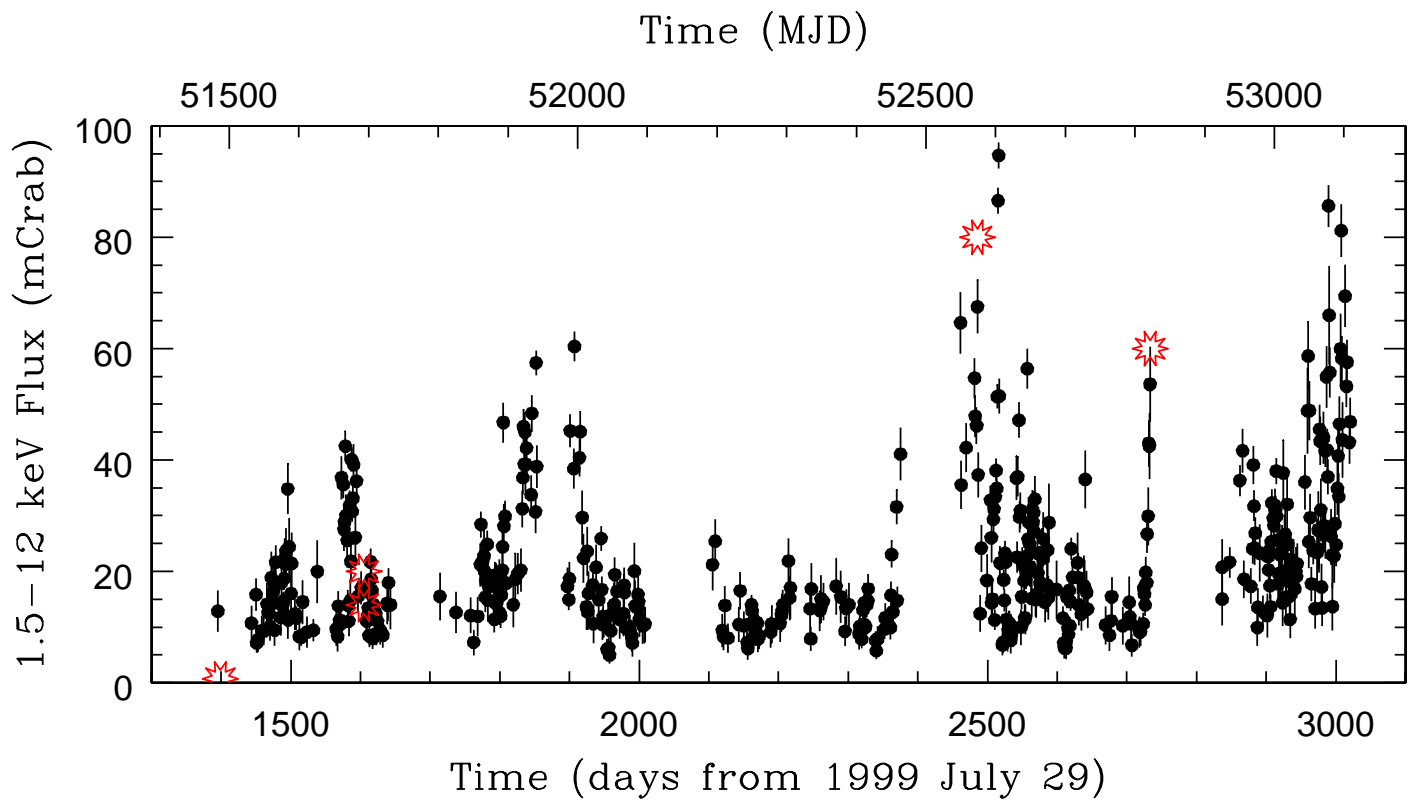}
\vspace{0in}\caption[h]{\footnotesize 1.5-12 keV {\em Rossi}-XTE ASM 
lightcurve of Mkn~421, from 1999, July 29 to 2004, April 14. 
Superimposed starred points are 0.5-2 keV {\em Chandra} flux measurements 
extrapolated to the 1.5-12 keV ASM band by using the best fitting Chandra 
continua.}
\end{figure}
\begin{figure}
\hspace{0in}
\epsfbox{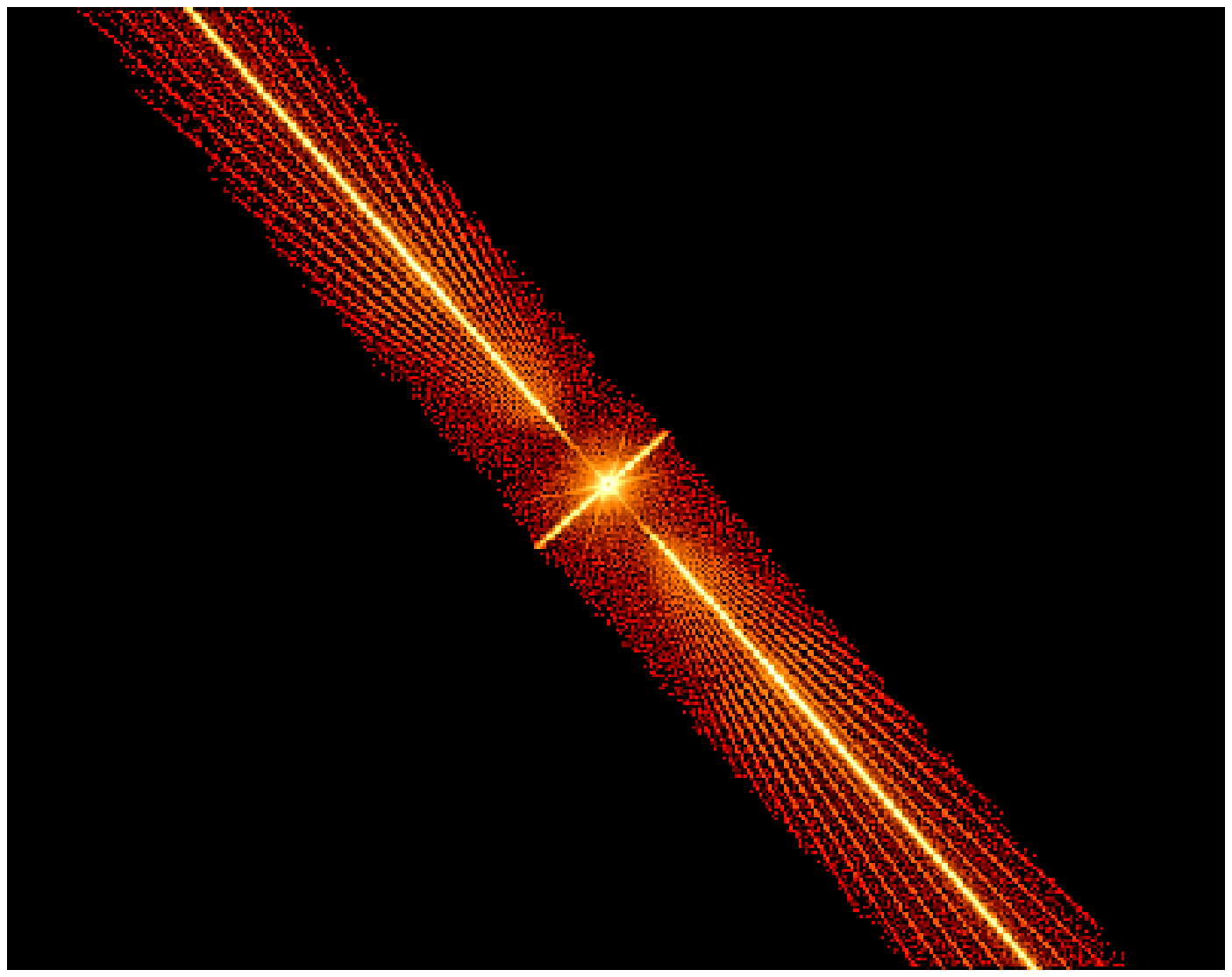}
\vspace{0in}\caption[h]{\footnotesize Portion of the 2-D ACIS-LETG dispersed 
spectrum of Mkn~421 during our first {\em Chandra} TOO on 2002, October, 
26-27. The dispersed spectrum is seen in the two very bright lines which 
form diagonal from the upper left to the bottom right, the positive and 
negative orders, and this is the highest signal-to-noise spectrum taken so 
far with the LETG . 
The ``fan of rays'' around both the 0th-order image and the 
dispersed spectrum are an instrumental artifact due to diffraction of the 
X-ray photons off the instrument support. The lack of photons (i.e. the 
``hole'') in the very center of the 0th-order source image is due to an 
exceptionally high degree of ``pile-up'' in the CCD detector. Finally 
the luminous strip centered on the 0th-order image perpendicular to the 
dispersed spectrum contains 0th-order photons 
deposited during read-out along the CCD read-out axis. This is because the source 
photon arrival rate is much faster than the 0.8 s CCD read-out rate.}
\end{figure}

\section{The Combined LETG Spectrum of Mkn~421} 
To increase the signal to noise, we combined the first order ACIS-LETG 
spectrum (and responses) of Mkn~421, with the two non-order-sorted HRC-LETG 
spectra (and responses) of the target in the overlapping 10-60 \AA\ 
wavelength range. The final co-added LETG spectrum contains more than 
7 million counts, and has $\sim 5300$ CPREs in the 
continuum at 21 \AA. 
This is the best S/N of any high resolution spectrum taken with 
{\em Chandra}, including Galactic X-ray binaries. 
This S/N was enough to detect OVII columns of $N_{OVII} \gs 8 \times 
10^{14}$ cm$^{-2}$ at a significance level $\ge 3\sigma$. 

\subsection{Spectral Responses and Continuum Fitting}
ACIS-LETG negative and positive 1st order effective areas (ARFs) were 
built by using the {\em 'fullgarf'} script distributed with {\em Ciao} 
(vs. 3.0.2), with the ACIS-Quantum-Efficiency (QE) file made 
available by the ACIS calibration group (CALDB 2.2.6) to correct for the 
ACIS contamination problem
\footnote{http://asc.harvard.edu/ciao/why/acisqedeg.html}
(e.g. Marshall, et al., 2003). Negative and 
positive ACIS-LETG 1st order ARFs were then checked for consistency and 
combined using the {\em Ciao} tool {\em add\_grating\_orders}, and finally 
convolved with the standard 1st order ACIS-LETG Redistribution Matrix (RMF, 
CALDB 2.2.6). 
For the HRC-LETG responses, instead, we used the standard convolution 
products of RMFs and ARFs files, for orders 1 through 6 
\footnote{http://asc.harvard.edu/cal/Letg/Hrc\_QE/ea\_index.html\#rsps}
. 
Finally, the first-order ACIS-LETG and order 1 through 6 HRC-LETG responses 
were combined together (weighting by count rate) to form 
a ACIS-LETG+HRC-LETG response matrix. 

\medskip
We used the fitting package {\em Sherpa} 
(Freeman et al., 2001), in 
{\em Ciao}, to model the broad band 10-60 \AA\
\footnote{We also excluded the wavelength range 48-57 \AA, where 
the two HRC plate gaps are present, and non modeled in the effective area} 
average continuum of Mkn~421 during the three LETG observations with a 
single power law absorbed by neutral Interstellar matter with solar 
composition (Morrison \& McCammon, 1983) along the line 
of sight. The best fitting hydrogen column of 
$N_H = (1.41 \pm 0.01) \times 10^{20}$ cm$^{-2}$ is in excellent agreement 
with the Galactic value along this line of sight ($N_H^{Gal} = 1.4 
\times 10^{20}$ cm$^{-2}$: 
Elvis, Lockman \& Wilkes, 1989; Dickey \& Lockman, 1990), and the 
best fitting power law has a slope of $\Gamma = 1.802 \pm 0.003$. This 
model, however, does not yield an acceptable fit to the data 
($\chi^2_r(d.o.f.) = 3.8(1635)$). While the large scale continuum shape is 
well accounted for, there are instead highly significant small scale 
systematic positive and negative residuals across the entire wavelength 
band (Figure 3). 
Hence attempts to reduce the differences between the 
model and the data by the standard method of introducing an additional 
continuum component, as a second power law, a broken power law, or a 
black-body, do not yield satisfactory results. This is due to the localized 
nature of the strong deviations, even the largest of which (at $\sim 43$ \AA)
would fit into a single resolution element of earlier CCD spectra. 

\subsubsection{Isolating Instrumental Artifacts in the Spectrum}
The strongest negative and positive features in the ratio 
between data and best fitting model, fall close to the neutral 
C and O K-edges respectively, at $\sim 23$ \AA\ and $\sim 43$ \AA\ (Fig. 3), 
as well as some of the ACIS-LETG bad columns associated to node-boundaries
\footnote{ACIS Node-boundaries (3 per chip) are normally 
associated with bad columns. These columns, however, are not completely 
blind in wavelength scale, due to the satellite dithering, and appear 
as smooth and broad ($\sim 1$ \AA) notch-like features in the 
aspect-reconstructed dispersed spectrum.}
or chip-gaps (particularly, the ones at $\sim 20$ \AA, and $\sim 13$ \AA, 
Fig. 3). 
%
\begin{figure}
\hspace{-0.8in}
\epsfbox{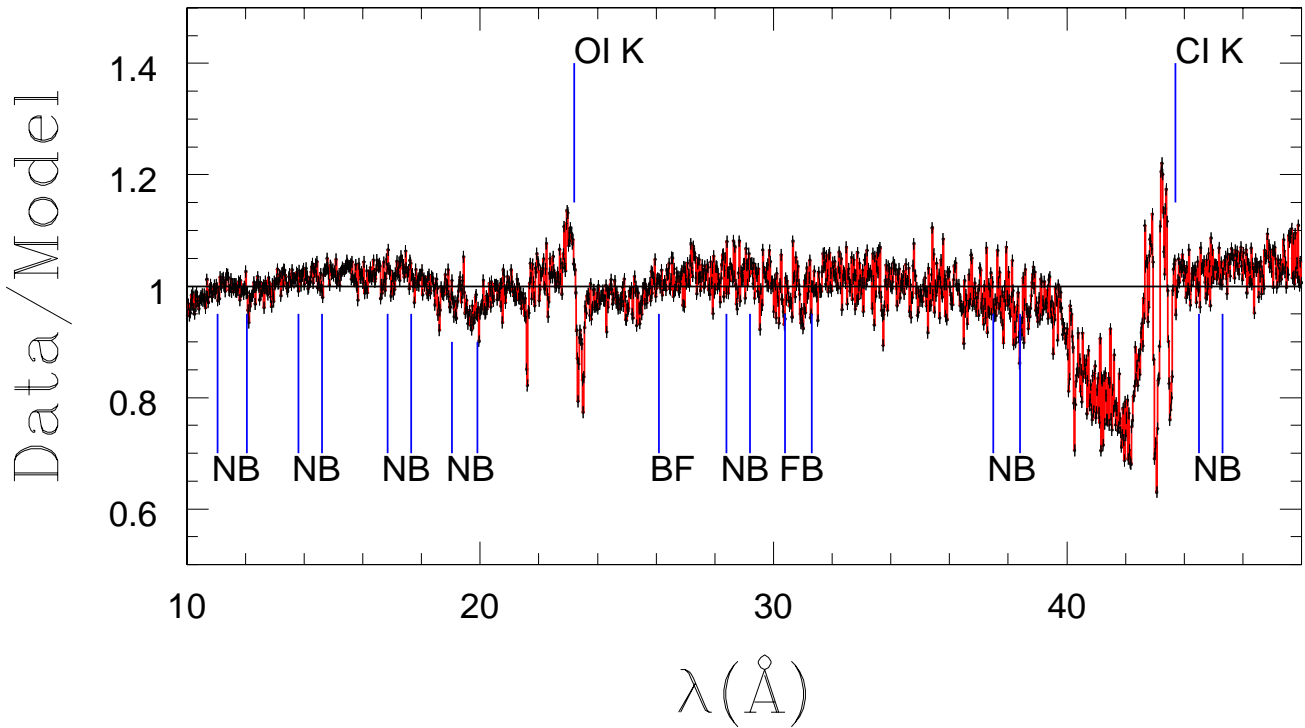}
\vspace{0in}\caption[h]{\footnotesize Ratio between the combined LETG spectrum of 
MKN~421 and its best fitting simple power law absorbed by neutral ISM with 
metal abundances fixed to solar value. NB = 'Node Boundary'; FB = 
'Front-Illuminated to Back-Illuminated'; BF = 'Back Illuminated to 
Front-Illuminated'.}
\end{figure}
%
All these features are most likely due to residual calibration 
uncertainties both in the ACIS- and HRC-LETG. 
In particular residuals from the CI and OI K-edge wavelengths shortward are 
likely to be mostly due to uncertainties in the model description of the 
instrumental absorption caused by a layer of contaminant material that has 
gradually condensed on the ACIS window, and that affects all observations 
taken with the ACIS detector (Marshall et al., 2003)
\footnote{http://cxc.harvard.edu/cal/Acis/Cal\_prods/qeDeg/}
. Clearly the depth of the CI edge is severely underestimated by the model 
(amplitude of the data/model ratio $\gs -30$ \%, Fig. 3), while the OI edge 
depth is overestimated (amplitude $\ls +10$ \%, Fig. 3). The exact shape 
of the recovery of the continuum level shortward of these two edges is 
important in setting the proper level of the local continuum, 
and so to properly evaluating the significance and amplitude of weak 
narrow absorption features in these spectral regions. 
As an example, Figure 4a shows the 20-25 \AA\ region of the LETG spectrum 
of Mkn~421, with the best fitting powerlaw plus neutral absorption continuum 
superimposed. The fit is unacceptable, with a reduced-$\chi^2$ of 
$\chi^2_r(d.o.f.) = 6.98(197)$. Clearly the OI K-edge is overestimated by 
the model, and so is all of the continuum between $\sim 21.7$ and 25 \AA, 
leaving strong systematic residuals in the data/model ratio. The strong 
narrow absorption features in the spectrum are OVII and OI absorption at 
$z=0$ (Fig. 4a). 
We also note here the scatter of the data in the narrow portion of 
the spectrum immediately longward of the OVII K$\alpha$ line at $z=0$ and 
up to $\sim 22.3$ \AA\ (where the OVII K$\alpha$ transitions at the redshift 
of Mkn~421 would fall), which may suggest the presence of additional 
narrow line-like features. 
Figure 4b shows the highly non-gaussian (60 bins outside the (-2,+2)$\sigma$ 
interval, versus the expected 7.5 in the gaussian hypothesis), 
positive-skewed (85 bins fall in the (0,+3)$\sigma$ interval, 
versus the 55 bins in the corresponding negative interval) and 
apparently bimodal (reflecting the systematics in the data/model ratio) 
distribution of the residuals in $\sigma$ in relative number of bins per 
half-$\sigma$, after ignoring narrow intervals of data at the wavelengths 
of the strong $z=0$ OVII, atomic OI and molecular OI absorption lines. 
%
\begin{figure}
\hspace{-0.8in}
\epsfbox{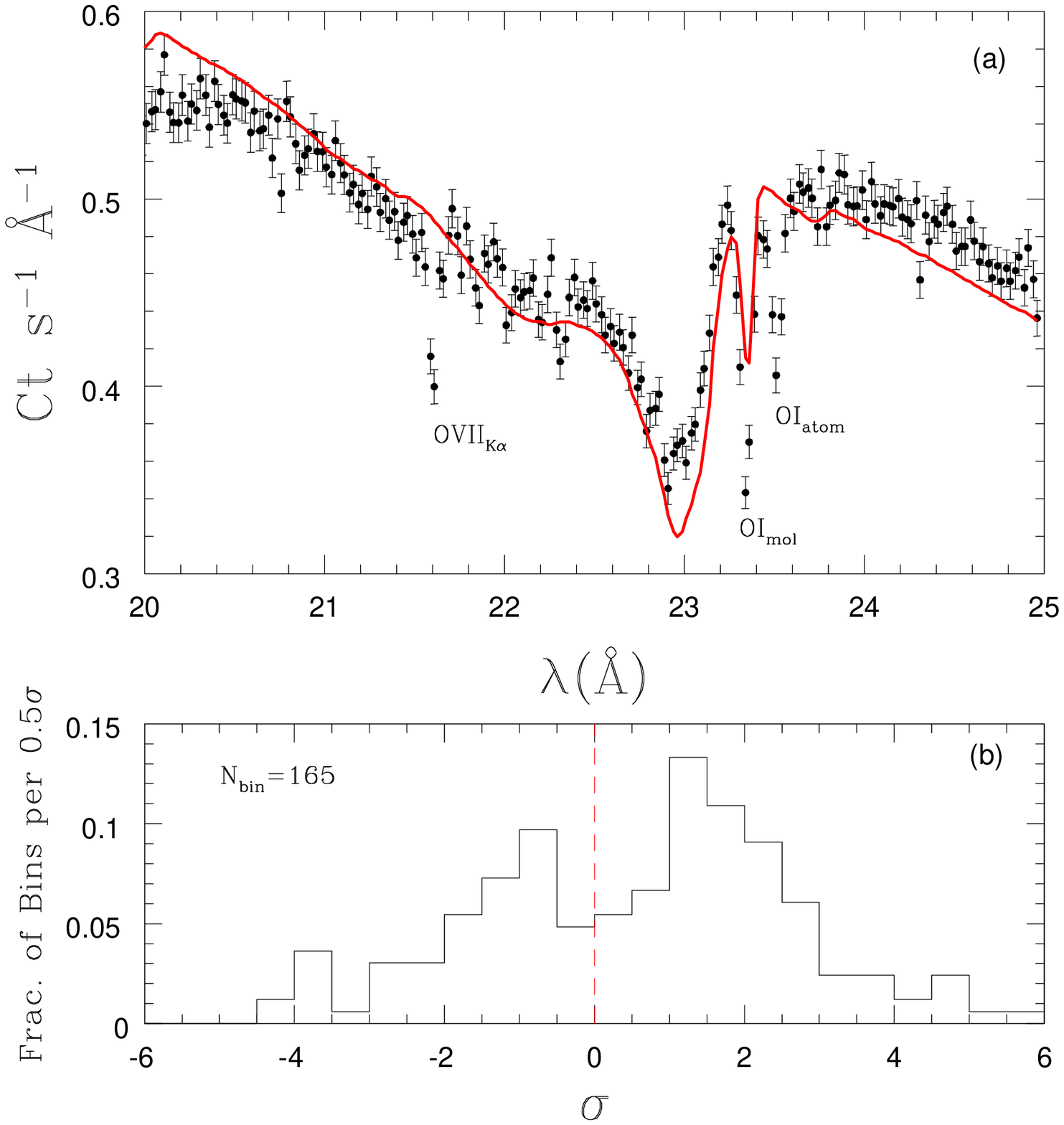}
\vspace{0in}\caption[h]{\footnotesize (a) 20-25 \AA\ region of the LETG 
spectrum of Mkn~421 and best fitting powerlaw plus neutral absorption 
continuum (solid curve). (b) Distribution of the fraction of residuals per 
half-$\sigma$ with $\sigma$.}
\end{figure}
%

We thus decided to re-fit the broad-band data leaving the metal abundances 
of the photoelectric absorption model used to model the ISM 
absorption, free to vary. 
This fit gave a best fitting H absorbing column of $(1.44 \pm 
0.01) \times 10^{20}$ cm$^{-2}$ (again in agreement with the Galactic ISM column observed in 
neutral H, Elvis, Lockman \& Wilkes, 1989), with C and Ne over-abundance of 
3 and 4 times solar respectively, and N and O virtually absent. 
We stress that these values should not be thought of as true differences 
in the ISM metallicity, compared to solar, but rather as due to calibration 
uncertainties. 
The best fitting power law slope now becomes $\sim 0.1$ steeper, $\Gamma = 1.911 
\pm 0.003$.  
The continuum model yields much better descriptions of 
the CI and OI K-edges (Fig. 5a,b), but it slightly amplifies the 
discrepancies between data and model at the wavelengths of several ACIS 
node-boundaries or chip-gaps, particularly around 13 and 20 \AA\ (Fig. 6a,b). 
Moreover locally the continuum level shortward of the CI and OI K-edges 
continues to be somewhat over/under-estimated (Fig. 5a,b). 
This may at least partly 
reflect the fact that the photoelectric model we use to describe the ISM 
absorption uses photoelectric cross-sections that do not properly 
describe the complicated absorption structure longward of the instrumental 
edge wavelengths
\footnote{http://space.mit.edu/CXC/calib/letg\_acis/ck\_cal.html}
. 
%
\begin{figure}
\hspace{-0.8in}
\epsfbox{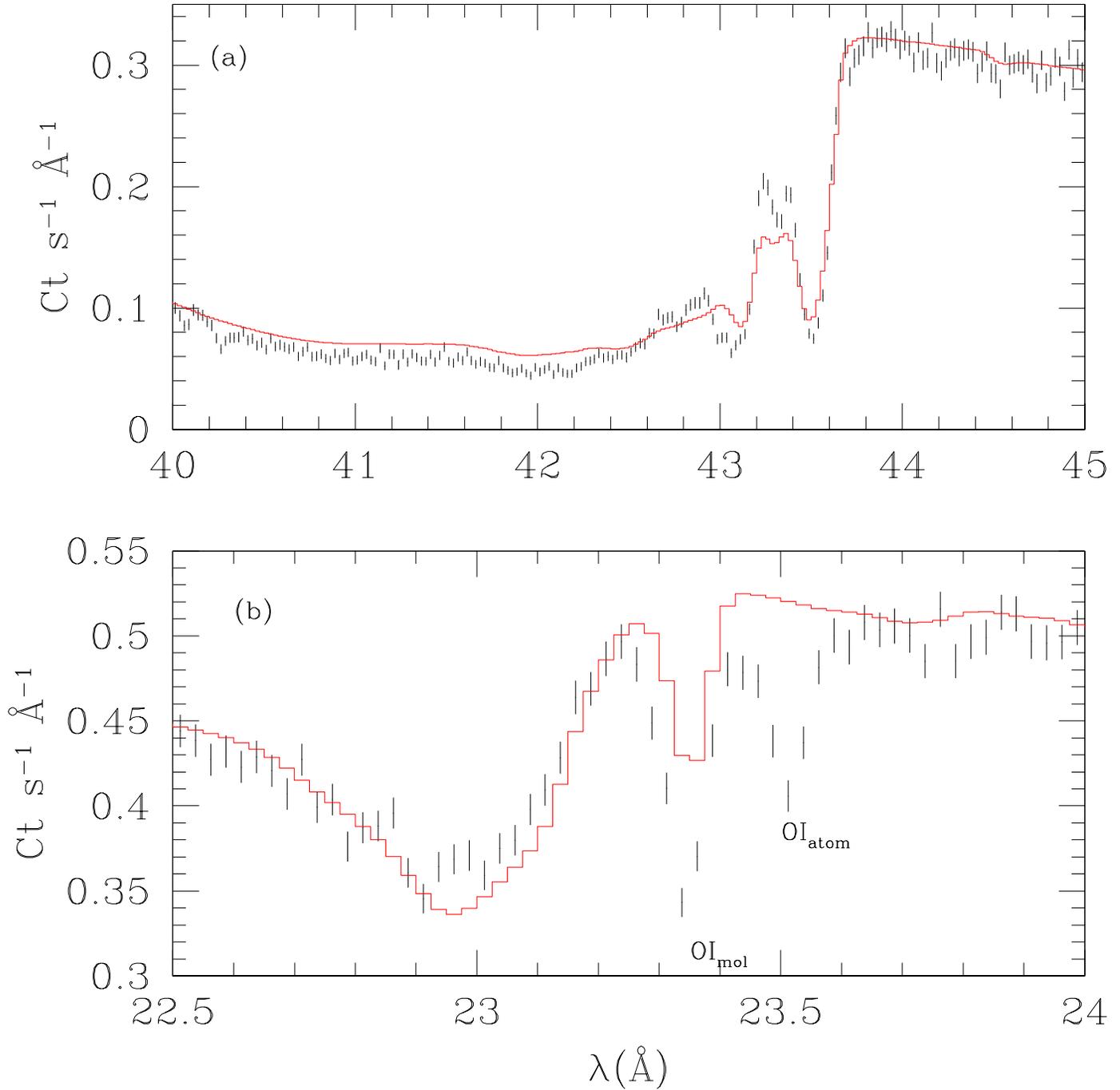}
\vspace{0in}\caption[h]{\footnotesize LETG spectrum of Mkn~421 and best 
fitting powerlaw plus neutral absorption with varying metallicities 
continuum (solid curves) around (a) the CI K-edge and (b) the OI K-edge.}
\end{figure}
%
%
\begin{figure}
\hspace{-0.8in}
\epsfbox{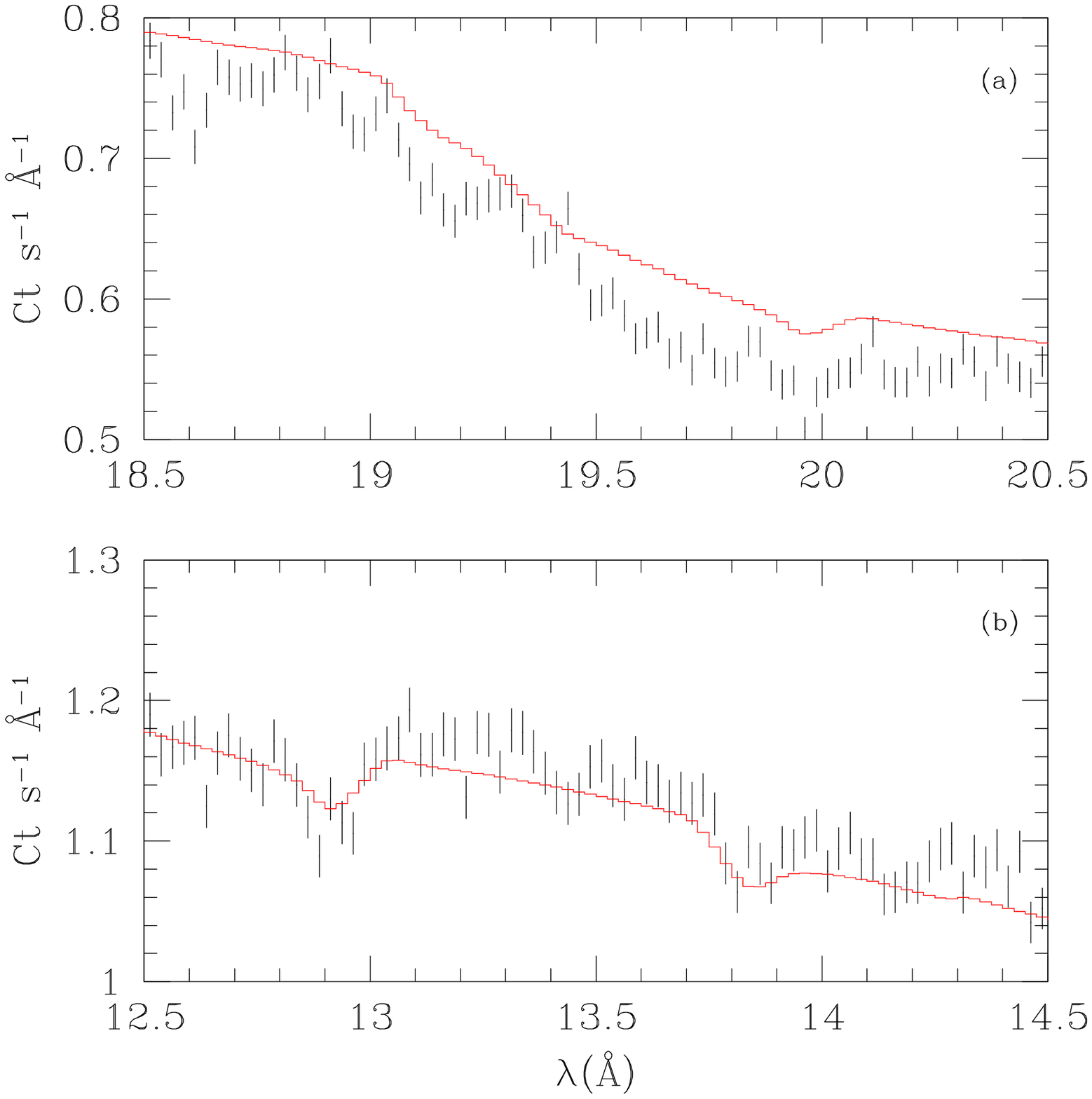}
\vspace{0in}\caption[h]{\footnotesize LETG spectrum of Mkn~421 and best 
fitting powerlaw plus neutral absorption with varying metallicities 
continuum (solid curves) around (a) 19-20 \AA\ and (b) 12.8-13.8 \AA\ 
ACIS node-boundaries.}
\end{figure}
%
To allow for these deviations we added ten broad gaussians (6 in emission 
and 4 in absorption). These have best fitting FWHM of 0.5-7 \AA, except at 
the exact wavelength of the CI K-edge where a relatively narrow gaussian, with 
FWHM of 0.2 \AA\ fits well. 
With this adjustment we recover a good fit to the continuum level in three 
important regions of the spectrum longward of the CI edge (between 
$\sim 40-43$ \AA, $\sim 26-30$ \AA\ and $\sim 31.5-35.5$ \AA), and at the 
wavelengths of two ACIS 
node-boundaries ($\sim 12-14$ \AA, and $\sim 19-21$ \AA). 
This model yielded a $\chi^2_r(dof) = 1.81(1598)$, and left unchanged 
both $\Gamma$ and $N_H$. 
Fig. 7a shows the residuals, in $\sigma$, between the data and our 
final best fitting continuum model. 
The residuals are now relatively flat over the entire 10-48 \AA\ 
wavelength range (Fig. 7a). There are still relatively 
broad excesses at the OI and CI K-edges wavelengths. There are also several 
narrow negative spikes, which we show in the next section, all due to intervening 
narrow resonant absorption metal lines. 
Fig. 7b show the normalized distribution of residuals to our best 
fitting continuum model, compared with a gaussians distribution. 
The positive tail of the distribution is gaussian. 
The negative-only $\le -3\sigma$ tail in the distribution is 
due to the presence of several real and high significance absorption 
lines in the data. 
%
\begin{figure}
\hspace{-0.8in}
\epsfbox{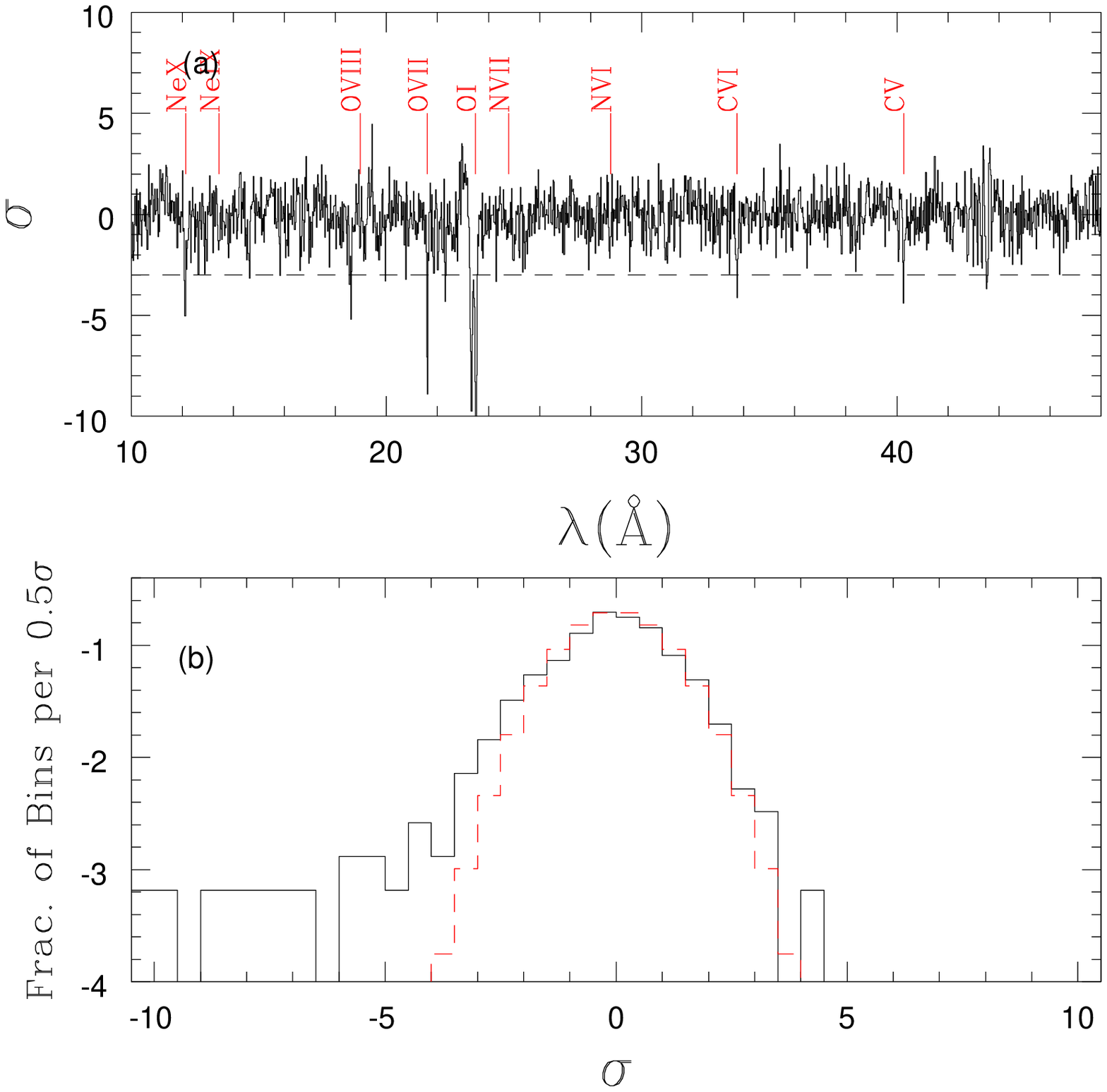}
\vspace{0in}\caption[h]{\footnotesize (a) Residuals in $\sigma$ of 
the LETG spectrum of MKN~421 to its best fitting continuum model: a power 
law absorbed by neutral ISM with metal abundances free to vary, plus ten 
broad gaussians used to recover the continuum level in spectral regions 
still affected by non-optimally modeled instrumental features. Rest Frame 
wavelengths of the K$\alpha$ transitions from He-like and H-like C, N, O 
and from neutral O, are labeled. 
(b) Normalized distribution of residuals to the best fitting continuum 
model described above (solid histogram), compared with the gaussian 
distribution (dashed histogram).}
\end{figure}
%

\subsection{Intervening Resonant Absorption Lines}
Our statistical analysis (Figure 7b) shows that an excess, compared to a 
gaussian distribution, of 28 $\ge 3\sigma$ negative residuals, 
with 16 of these being $\ge 4\sigma$, are present in the LETG spectrum of 
Mkn~421, relative to the best fitting continuum model. Inspection of 
relatively narrow regions ($\Delta \lambda = 2-5$ \AA) of the spectrum 
(Fig. 8a,f), confirms these findings and shows that several groups of 2-4 
contiguous 25 m\AA\ negative bins deviate from the best fitting continuum 
with significance $\ge 1.5\sigma$ per bin (i.e. $\ge 3\sigma$ per group). 
It is striking that these narrow absorption line-like features concentrate 
in critical spectral regions, always at, or longward of, the rest-frame 
positions of the main resonant transitions from highly ionized or 
neutral C, N, O and Ne (Fig. 8). 
%
\begin{figure}
\hspace{-0.8in}
\epsfbox{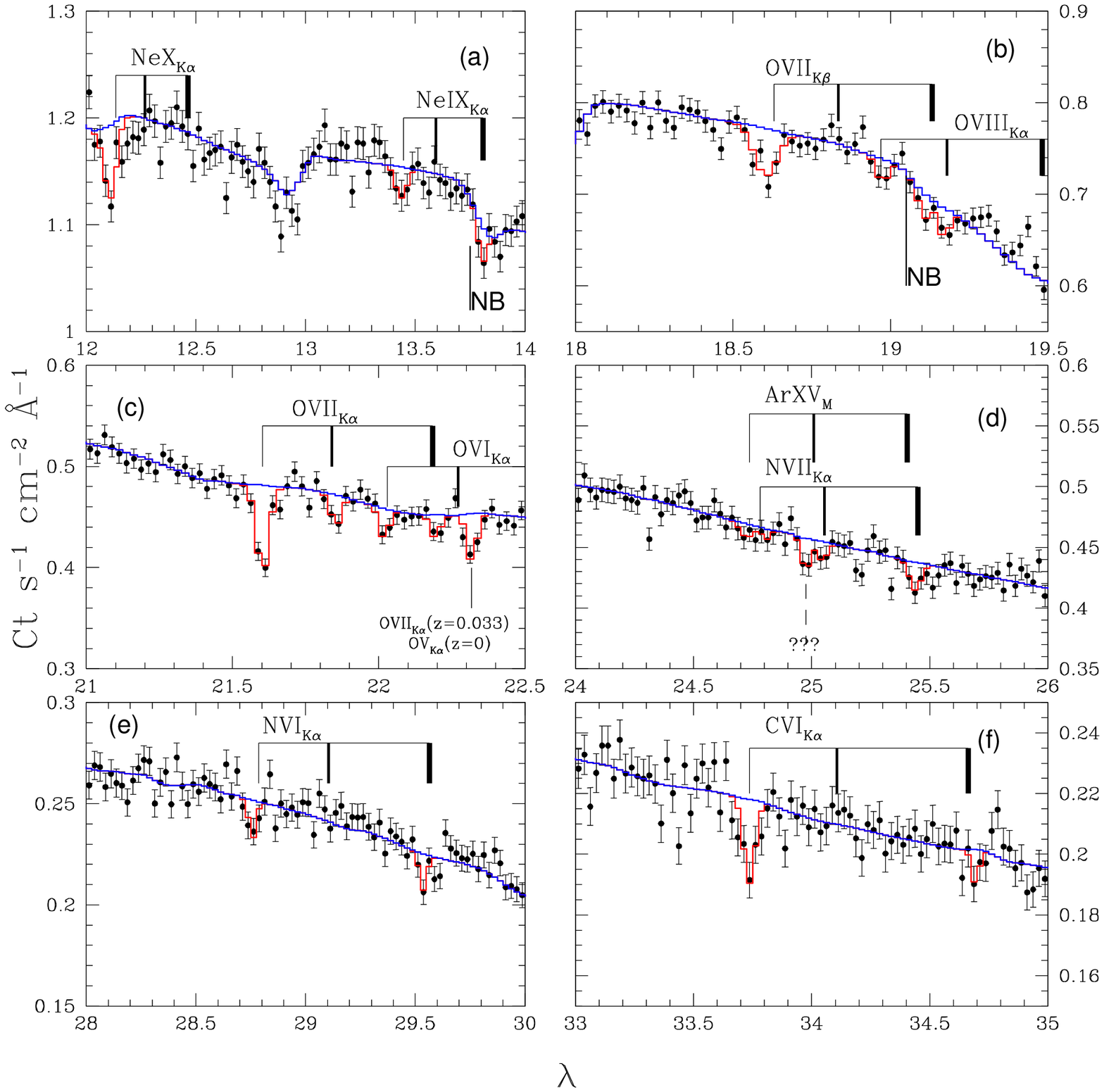}
\vspace{0in}\caption[h]{\footnotesize Six portions of the LETG spectrum of 
Mkn~421 along with it best fitting continuum plus narrow absorption model 
(solid line), centered around the rest-wavelengths of the (a) 
$NeXI-X_{K\alpha}$, (b) OVIII$_{K\alpha}$, (c) OVII$_{K\alpha}$, 
(d) NVII$_{K\alpha}$, (e) NVI$_{K\alpha}$ and (f) CVI$_{K\alpha}$ 
transitions. For each labeled ion, the three vertical lines from left to 
right are the rest frame wavelength (thin line) and the expected wavelengths 
at $z=0.011$ (medium thickness line) and $z=0.027$ (thick line).}
\end{figure}
%
We therefore tentatively identified these features 
as metal absorption lines from either the interstellar medium (ISM) of our 
Galaxy, the Local group WHIM Filament (Nicastro et al., 2002, 
Nicastro et al., 2003, Williams et al., 2005), intervening intergalactic 
absorbers, or 
material outflowing from the nucleus of Mkn~421 along our line of sight. 
We fitted these features with negative gaussians, leaving all parameters 
(FWHM, centroid and amplitude) free to vary. 
Our best fitting model thus contains a total of 24 unresolved (i.e. intrinsic 
FWHM $< 50$ m\AA) absorption lines, and has a $\chi^2_r(dof) = 1.25(1529)$ 
(giving an F value of 14.2, when compared with the best fitting continuum 
model with no lines, corresponding to a probability of exceeding F, P$(>F) << 
0.001$). 
Figure 9 shows the normalized distribution of residuals to our best 
fitting model, including absorption lines, over the 10-48 \AA\ range, compared 
with a gaussian distribution. 
The negative tail present in the distribution of Fig. 7b, 
is no longer visible, and the overall distribution is now close 
to gaussian. 
We are therefore confident that the measured significance of 
our absorption lines, as expressed in number of standard deviations from a 
parent gaussian distribution, represents a good estimate of the significance 
of the lines. 
%
\begin{figure}
\hspace{-0.8in}
\epsfbox{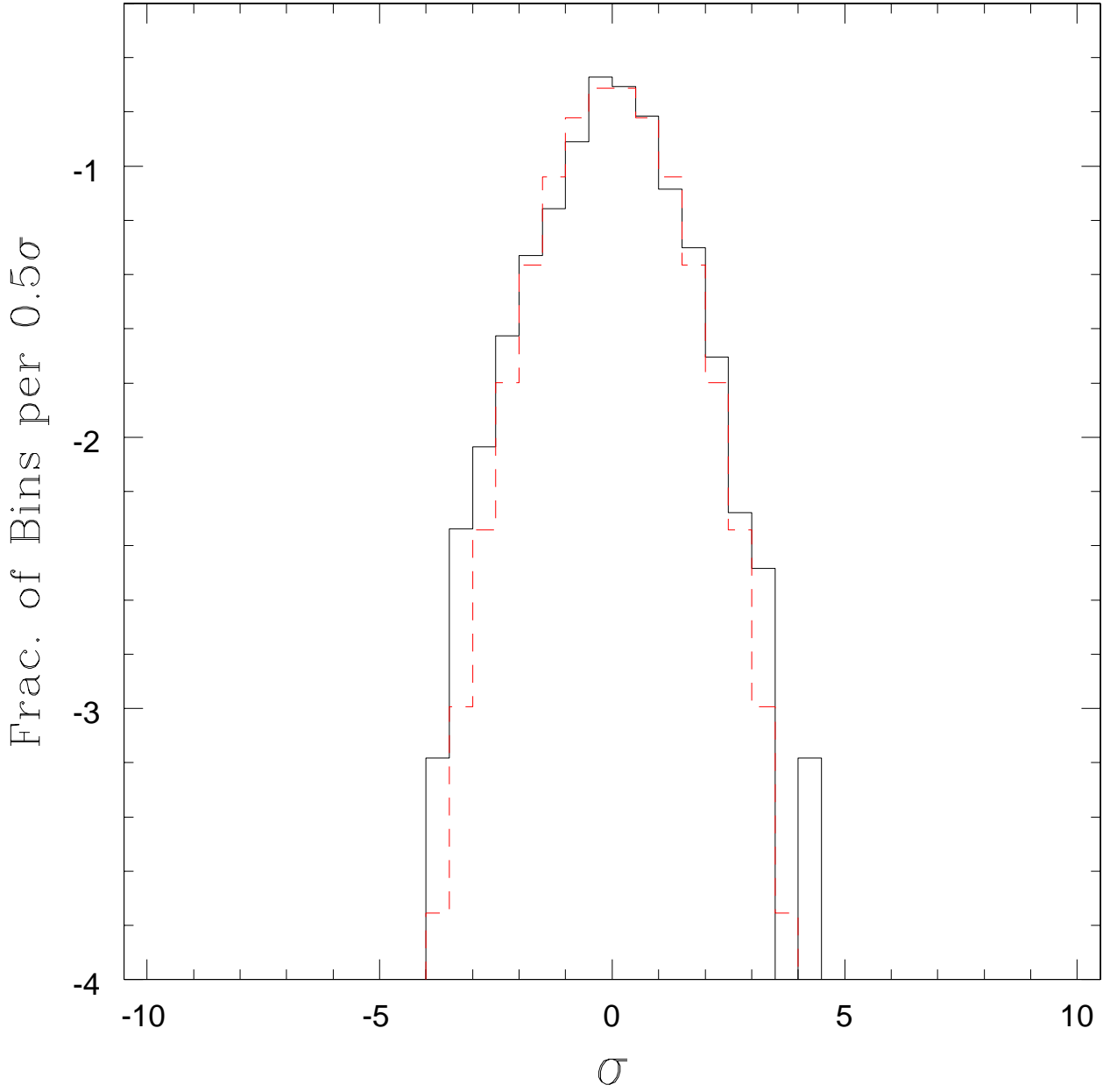}
\vspace{0in}\caption[h]{\footnotesize Normalized distribution of residuals 
to our best fitting model, including 24 narrow (FWHM$< 50$ m\AA) 
absorption gaussians. The dashed histogram is the gaussian distribution.}
\end{figure}
%

Figure 8 (a to f) shows 21 of the 24 absorption lines fitted
\footnote{Not shown are the neutral O from the ISM, and the CV K$\alpha$ 
at $z=0$ (Williams et al., 2005).} 
in 6 different 
portions of the LETG spectrum of Mkn~421, centered around the 
rest-wavelengths of the NeXI-X$_{K\alpha}$, OVIII$_{K\alpha}$, 
OVII$_{K\alpha}$, NVII$_{K\alpha}$, NVI$_{K\alpha}$ 
and CVI$_{K\alpha}$ transitions. 

\subsubsection{Line Identification}
We identify 14 of the 24 absorption lines in the LETG spectrum of Mkn~421 as  
being produced by $z = 0 \pm 0.001$ ($cz = \pm 300$ km s$^{-1}$) systems of 
neutral (molecular and atomic OI) and ionized metals. Physically the system 
corresponds to two distinct components: the ISM of the Milky 
Way (i.e. Savage et al., 2000, Savage et al., 2003, Savage et al., 2005, ApJ) 
and the Local Group WHIM (e.g. Nicastro et al., 2002; 
Nicastro, 2003, and references therein). 
One more line, at $\lambda = (22.32 \pm 0.02)$ \AA\ (Figure 8c) is 
tentatively 
identified here with the combination of the $z=0$ OV K$\alpha$ inner 
shell transition ($\lambda_{rest-frame} = 22.37$ \AA\, Schmidt et al., 
2004) and an OVII K$\alpha$ 
transition at $z=0.033 \pm 0.001$ ($\sim (900 \pm 300)$ km s$^{-1}$ bluer 
than the estimated redshift of the blazar's host galaxy: De Vacouleurs, 
1991). 
These local systems are the subject of a forthcoming paper (Williams et al., 
2005), and will not be discussed further here. 

The remaining 10 lines could not be identified with any other $z=0$ 
transition from either neutral or ionized metals. 
Two lines ($\lambda = 13.80$ \AA\ and $\lambda = 25.04$ \AA, see Table 2) 
fall close to the rest frame wavelengths of two outer shell transitions 
from FeXVII, at $\lambda=13.823$ \AA, and ArXII at $\lambda=25.04$ \AA\ 
(Verner, Barthel \& Tytler, 1994). 
However, these two transitions have oscillator strengths [$f(FeXVII) = 
0.331$ and $f(ArXII) = 0.191$: Verner, Barthel \& Tytler, 1994] that are 
respectively, 8.9 
and 7.6 times weaker than the oscillator strengths of two other transitions 
from the same ions: the FeXVII at $\lambda=15.01$ \AA\ ($f=2.95$) and the 
ArXII doublet at $<\lambda> = 31.37$ \AA\ ($f_1+f_2=1.45$). 
These are two regions of the 
spectrum not directly contaminated by known instrumental features and with 
effective areas comparable, or slightly lower, than the effective areas at 
the wavelengths of the detected features. So both the 
FeXVII$_{\lambda 15.01}$ and the ArXII$_{\lambda31.37}$ lines should be 
detected in this spectrum at a much larger significance than that of the 
putative $z=0$ lines at 13.80 and 25.04 \AA\ respectively. No line is 
instead detected at these wavelengths so ruling out the $z=0$ FeXVII and 
ArXII identifications for these two features. 
Finally, although the vast majority of the inner shell, X-ray, transitions 
from C, N O and Ne are poorly known, both in wavelength and strength (ATOMDB, 
v.1.3.0, Smith et al., 2001; Behar \& Netzer, 2002), basic atomic physics 
requires that the strongest of these transitions (K$\alpha$) must all fall 
at longer wavelengths than the K$\alpha$ He-like transitions from the same 
elements. 

We detect 5 lines in the OVII K$\alpha$ $z=0-0.033$ range (Figure 8c). 
From the left, the first line is unambiguously identified with the 
OVII K$\alpha$ at $z=0$. The next line to the right, at $\lambda = 21.85$ \AA, 
cannot be due to inner-shell absorption by the lower-ionization ion of oxygen, 
OVI, since the strongest transition from this ion ($1s^22s \rightarrow 
1s2s2p$) falls at $\lambda = 22.03$ \AA\ (HULLAC value: Bar-Shalom et al., 
2001; Liedhal, D.A., private communication), that is redward of the line 
at $\lambda = 21.85$ 
\AA\ (moreover the OVI K$\alpha$ line at $z=0$ is indeed detected in our 
spectrum, as labeled in Figure 8c; see also Williams et 
al., 2005). Similarly, the next line to the right, at $\lambda = 22.20$ 
\AA, cannot be due to OV absorption, since the strongest 
inner-shell transition from this ion falls at $\lambda = 22.37$ \AA\ (Schimdt 
et al., 2004), that is redward of the line at $\lambda = 22.20$ \AA. A 
weak OV K$\alpha$ inner-shell transition may indeed be detected in our 
spectrum, in the red wing of the line at $\lambda = 22.32$ \AA (Figure 8c). 
Finally, three strong (i.e. oscillator strength $\ge 0.07$) inner-shell 
transitions from the next lower ionization ion, OIV, have rest frame 
wavelengths $\lambda = 22.73-22.75$ \AA (unresolved in the LETG: HULLAC 
values, Liedhal, D.A., private communication). These 
wavelengths are longward of the OVII K$\alpha$ wavelength at the redshift 
of Mkn~421. We conclude that the two weak lines at $\lambda =21.85$ and 
$\lambda = 22.20$ \AA\ must be due to highly ionized intervening O 
absorption at $z>0$. 

We repeated this analysis for the other three portions of the spectrum 
of Mkn~421 that maybe in principle affected by inner-shell C, N and Ne 
absorption, redward of the K$\alpha$ transitions from the He-like 
ions of these elements. 
We concluded that only two of the remaining 8 features may in principle 
be due to inner shell transitions from moderately ionized Ne or N: 
the line at $\lambda = 13.80$ \AA\ (Figure 8a) and the line at 
$\lambda = 29.54$ \AA\ (Figure 8e; see Table 2). The line at 
$\lambda = 13.80$ \AA\ lies close to the rest frame wavelength of the 
forbidden line of the NeIX triplet, analogous to the OVI 
K$\alpha$ inner shell transition that lies between the intercombination 
and the forbidden lines of the OVII triplet. 
We then hypothesized that the $\lambda = 13.80$ \AA\ line be due to NeVIII 
K$\alpha$ inner shell absorption. Scaling by wavelength ratio from the 
'OVII-triplet plus OVI-K$\alpha$' system we derived a wavelength for the 
NeVIII K$\alpha$ of $\lambda = 13.78$ \AA, consistent with the observed one 
within the $\Delta\lambda =20$ m\AA\ calibration uncertainties (see below). 
Similarly, we tried to estimate the wavelength of the NV K$\alpha$ inner 
shell transition scaling from the OVII+OVI system, and found 
$\lambda(NV_{K\alpha}) = 29.44$ \AA, in good agreement with the HULLAC value 
of $\lambda = 29.46$ \AA. 
This is significantly different from 
the wavelength of the observed feature at $\lambda = 29.54$, ruling out the 
$z=0$ NV K$\alpha$ identification. We conclude that at most one of the ten 
detected lines, at $\lambda = 13.80$ \AA, can be tentatively identified with 
a $z=0$ line from NeVIII. 

All other lines must belong to redshifted systems. Nine of the 10 lines
\footnote{The tenth line is clearly present at an integrated significance 
of 5$\sigma$ at $\lambda = 24.97$ \AA, but could not be identified with any 
known transitions at the redshift of any of the three systems.}
can indeed be identified as belonging to 2 different intervening absorption 
systems, at $z = (0.011 \pm 0.001)$ [$cz = (3300 \pm 300)$ km s$^{-1}$] and 
$z = 0.027$ [$cz = (8090 \pm 300)$ km s$^{-1}$; Table 2]. We focus on these 
two systems in this paper.
The strongest line from the absorber at $z = 0.011 \pm 0.001$, is the 
OVII$_{K\alpha}$, while for the $z = 0.027 \pm 0.001$ absorber 
the two strongest lines are the NVII$_{K\alpha}$ and the NVI$_{K\alpha}$. 
The first five columns of table 2 list, in order of increasing wavelengths, 
measured line wavelengths, equivalent widths (EWs), ion column densities and 
significances for all lines for which a $z=0$ identification could not be 
found
\footnote{Included is the line at $\lambda = 13.80$ \AA, which we identify 
here with a NeIX K$\alpha$ line from the $z=0.011$ system, but that could 
be at least partly due to NVIII K$\alpha$ absorption at $z=0$. This does 
not affect our conclusion, however, since we consider the EW 
of this line as a 3$\sigma$ upper limit, due to the super-imposition of 
this feature with a CCD node-boundary.}
. 
The last three columns of Table 2, list the proposed line 
identifications, rest-frame wavelengths and redshifts. 

\subsubsection{Absorption Line Strengths}
We estimated the significance of the absorption lines by two 
methods: (1) conservatively (Table 2, Col. 4), by computing the $3\sigma$ 
error on their EWs, leaving all gaussians parameters 
and the continuum free to vary; 
(2) by directly integrating the deviations across the line profile (Table 2, 
Col. 5), from the fixed best fitting continuum. 
Direct integration in all cases gives larger significance. 
The significance of the lines depends critically on the degree 
of accuracy with which the local continuum is known. 
To assess this dependence for the 
the weakest lines in our spectrum we repeated locally the $\chi^2$ test 
on our best fitting continuum for all the 'interesting' regions plotted in 
Fig. 8, after eliminating narrow (50-100 m\AA) intervals of data 
centered on the $z=0$ lines. We let the normalization of our best fitting 
continuum vary locally, and re-fitted the data. In all cases, we obtained 
slightly lower relative normalizations (with value ranging from 0.95\% to 
0.98\%) and unacceptably high reduced $\chi^2$ values, for the given number 
of {\em d.o.f.}. 
As an example the 21.4-22.5 \AA\ interval 
gave a continuum relative normalization of 98.9\% and $\chi^2_r(d.o.f.) 
= 1.4(23)$, corresponding to a probability of exceeding $\chi^2$ of 
$P_{cont}(>\chi^2) = 9$ \%. 
Including gaussians for the two lines at 21.85 \AA\ and 22.20 
\AA, and re-fitting the data leaving again the relative continuum 
normalization and the 6 parameters of the gaussians free two vary, gives a 
relative continuum normalization of 1.0 and $\chi^2_r(d.o.f.) = 1.0(17)$, 
corresponding to $P_{cont+lines}(>\chi^2) = 45$ \%. An F-test between the 
two models, gives a probability of 10\% of exceeding F. All other cases gave 
lower $P_{cont}(>\chi^2)$ and $P(>F)$. We conclude that our estimate 
of the local continuum level in the regions where the absorption lines are 
present is statistically acceptable, and so confirm the goodness of the 
estimate of the significances of the lines shown in Table 2 (Col. 4 and 5). 

The NeIX$_{K\alpha}$ and OVII$_{K\beta}$ from the system at $z=0.027$, 
as well as the OVIII$_{K\alpha}$ line from the system at $z=0.011$, 
although clearly detected in the spectrum, lie close 
to ACIS detector chip node boundaries (Fig. 8a,b). 
These boundaries may induce an inaccurate estimate of the line EWs. 
We consider the EWs of these lines only as upper-limits. This does not 
affect our main conclusions on the physical state of the two absorbers, as 
shown in sections 7.3 and 7.4.. A 3$\sigma$ 
upper limit for the OVIII$_{K\alpha}$ from the $z=0.027$ system, which is not 
detected in the spectrum, is also listed in Table 2, since it will be 
used in our diagnostic analysis in \S 6.4. 

%
\begin{sidewaystable}
\footnotesize
\begin{center}
\caption{\bf \small Best fitting UV and X-ray absorption line parameters} 
\vspace{0.4truecm}
\begin{tabular}{|ccccc|ccc|}
\hline
& & & {\em Chandra}-LETG & & & & \\
\hline
$\lambda$ & EW$^a$ & Column & Conservative & Direct Integ. & Line Id. & 
$\lambda_{rf}$ & Redshift \\
          &        &        & significance & significance  &          &  
               &          \\
(\AA) & (m\AA) & $10^{15}$ cm$^{-2}$ & $\sigma$ & $\sigma$ & & \AA\ & \\
\hline
$13.80\pm 0.02$ & $< 1.5$$^c$ & $< 1.2$$^c$ & [3] & 3.8 & NeIX$_{K\alpha}$ & 
13.447 & $0.026 \pm 0.001$ \\
$19.11 \pm 0.02$ & $< 1.8$$^c$ & $< 3.8^c$ & [3] & 4.0 & OVII$_{K\beta}$ & 
18.63 & $0.026 \pm 0.001$ \\
$19.18\pm 0.02$ & $< 4.1$$^c$ & $< 3.0^c$ & [3] & 4.9 & OVIII$_{K\alpha}$ & 
18.97$^b$ & $0.011 \pm 0.001$ \\
19.477 (fixed) & $< 1.8$ & $< 1.3^c$ & [3] & [3] & OVIII$_{K\alpha}$ & 
18.97$^b$ & 0.027 (fixed) \\
$21.85 \pm 0.02$ & $3.0^{+0.9}_{-0.8}$ & $1.0^{+0.3}_{-0.3}$ & 3.8 & 6.3 & 
OVII$_{K\alpha}$ & 21.602 & $0.011 \pm 0.001$ \\
$22.20 \pm 0.02$ & $2.2 \pm 0.8$ & $0.7 \pm 0.3$ & 2.8 & 3.8 & 
OVII$_{K\alpha}$ & 21.602 & $0.028 \pm 0.001$ \\
$25.04 \pm 0.02$ & $1.8 \pm 0.9$ & $0.8 \pm 0.4$ & 2.0 & 3.2 & 
NVII$_{K\alpha}$ & 24.782$^b$ & $0.010 \pm 0.001$ \\
$25.44 \pm 0.02$ & $3.4 \pm 1.1$ & $1.4 \pm 0.5$ & 3.1 & 5.8 & 
NVII$_{K\alpha}$ & 24.782$^b$ & $0.027 \pm 0.001$ \\
$29.54 \pm 0.02$ & $3.6 \pm 1.2$ & $0.7 \pm 0.2$ & 3.0 & 4.6 & 
NVI$_{K\alpha}$ & 28.787 & $0.026 \pm 0.001$ \\
$34.69 \pm 0.02$ & $2.4 \pm 1.3$ & $0.5 \pm 0.3$ & 1.8 & 3.3 & 
CVI$_{K\alpha}$ & 33.736$^b$ & $0.028 \pm 0.001$ \\
\hline
& & & FUSE & & & & \\
\hline
1042.41 (fixed) & $< 18^c$ & $< 0.014^c$ & [3] & & OVI$_{2s\rightarrow 2p}$ & 
1031.926 & 0.01 (fixed) \\
1043.28 (fixed) & $< 21$ & $< 0.016$ & [3] & & 
OVI$_{2s\rightarrow 2p}$ & 1031.926 & 0.011 (fixed) \\
1059.79 (fixed) & $< 19^c$ & $< 0.014^c$ & [3] & & OVI$_{2s\rightarrow 2p}$ & 
1031.926 & 0.027 (fixed) \\
\hline
& & & HST-GHRS & & & & \\
\hline
$1228.02 \pm 0.01$ & $92.0 \pm 3.3$ & $0.0180 \pm 0.0006$ & 27.9 & 97.2 & 
HI$_{Ly\alpha}$ & 1215.67$^b$ & $0.010160 \pm 0.000007$ \\
1229.04 (fixed) & $< 26^c$ & $< 0.0047^c$ & [3] & & HI$_{Ly\alpha}$ & 
1215.67$^b$ & 0.011 (fixed) \\
1248.85 (fixed) & $< 47^c$ & $< 0.0085^c$ & [3] & & 
HI$_{Ly\alpha}$ & 1215.67$^b$ & 0.027 (fixed) \\
\hline 
\end{tabular}
\end{center}
$^a$ Absorption line EW are formally negative; here we drop the sign.
$^b$ Central wavelength of the two lines of the doubled, computed by
weighting their positions by their oscillator strengths. 
$^c$ 3$\sigma$ upper limit. 
\end{sidewaystable}
\normalsize
%

Uncertainties in the position of the line centroids are dominated 
by systematics in the LETG wavelength calibrations. Non-linearity in the 
LETG dispersion relationship produces an RMS deviation of predicted vs 
observed line positions of 13 m\AA\ (Drake, Chung, Kashyap, Ratzlaff, 2003), 
equivalent to 195 km s$^{-1}$ at 20 \AA. 
This uncertainty affects all spectra taken with the HRC-LETG configuration, 
but may be present at a lower extent also in spectra taken with the ACIS-LETG 
configuration. In our combined LETG spectrum, this uncertainty easily 
dominates the few tenths of m\AA\ 1$\sigma$ uncertainty in the gaussian 
centroid for the $z \simeq 0$ OVII$_{K\alpha}$ line (Table 2). 
We checked the amplitude of the LETG wavelengths accuracy by looking for 
differences between expected and observed relative positions of two strong 
lines from the same ion, in the same system. The largest discrepancy that 
we find is between the position of the OVII$_{K\alpha}$ and OVII$_{K\beta}$ 
at $z\simeq 0$, for which we measure a relative difference of $\Delta \lambda 
= 20$ m\AA\ for a line separation of $\sim 3$ \AA. For the 
absorption lines of the two $z>0$ systems, this is $\sim 5-50$ times larger 
than the statistical errors on the lines centroid. 
This calibration error fluctuates with wavelength, so that a simple shift 
is not sufficient to correct for it. 
We have then added this systematic uncertainty 
to all line centroids wavelength errors in our analysis. This translates into 
average redshift uncertainty $\sim \Delta z = 0.001$ (i.e. $\Delta (cz) \sim 
300$ km s$^{-1}$) at 20 \AA. 

Figures 10 and 11 show all the detected 
metal lines from our two intervening systems in velocity space as 
residuals (in $\sigma$) between the best fit continuum and the data. 
The vertical dashed line across the panels is the average redshift of the system. 
Because of the systematic LETG wavelength errors the line wavelengths have been 
shifted to the mean $z$ of the system. 
These difference are always within the estimated systematics in the 
wavelengths calibration (i.e. $\Delta \lambda < 20$ m\AA). 
%
\begin{figure}
\hspace{-0in}
\epsfbox{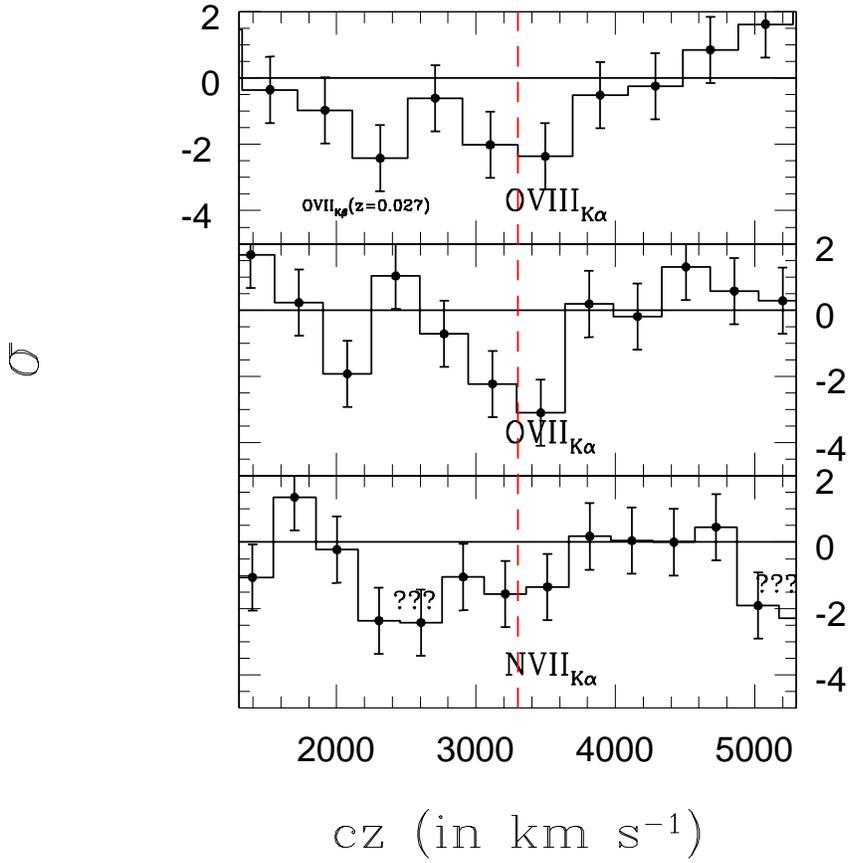}
\vspace{0in}\caption[h]{\footnotesize Residuals in $\sigma$, in velocity 
space, for three absorption lines of the $cz = 3300 \pm 
300$ km s$^{-1}$ system.}
\end{figure}
%

%
\begin{figure}
\hspace{+1in}
\epsfbox{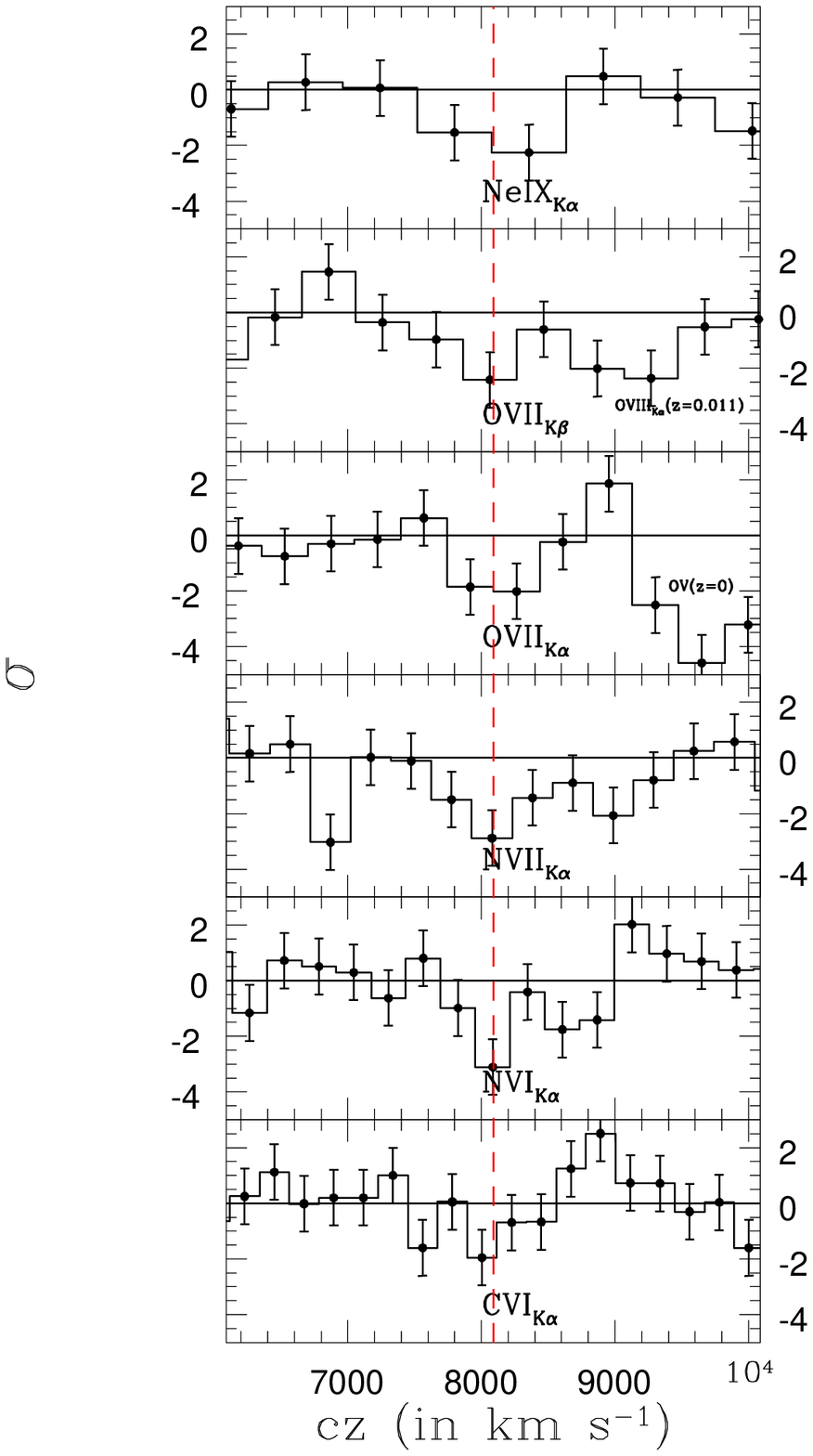}
\vspace{0in}\caption[h]{\footnotesize Residuals in $\sigma$, in velocity 
space, for six absorption lines of the $cz = 8090 \pm 
300$ km s$^{-1}$ system.}
\end{figure}
%

\subsubsection{Statistical Significance of the Two Absorption Systems}
We estimated the significance of the two systems at $z=0.011$ and $z=0.027$ 
by comparing the best-fit $\chi^2$ of simpler models that do not include the 
metal absorption lines from these systems with that of more complex models 
that include them. 
The F-tests were run over the following two portions of the spectrum, 
containing a similar number (100 and 130) of resolution elements: 
$\Delta\lambda(z=0.011) = [21,26]$ \AA\ and $\Delta\lambda(z=0.027) = 
[21.5,23]\bigcup[24.5,26]\bigcup[27.5,30]\bigcup[34,35]$ \AA. 
OVII$_{K\alpha}$ and NVII$_{K\alpha}$ lines were included in the complex 
model for the $z=0.011$ system, and OVII$_{K\alpha}$, NVII$_{K\alpha}$, 
NVI$_{K\alpha}$ and CVI$_{K\alpha}$ were used for the $z=0.027$ system. 
The fits with the complex models were performed freezing the width 
of the lines to the instrument resolution and fixing their relative 
positions and normalizations to their expected relative position based on 
the proposed identifications, and their measured relative normalization. 
Only the redshift of the system and the overall normalization were left free 
to vary in the fits. 
The results of these tests are presented in Table 3, in which we 
list both the probability of exceeding F (Col. 3) and the associated 
significance in $\sigma$ (Col. 4). 
The last column of Table 3, lists the sum of the lines significances 
for the two systems (Col. 4 of Table 2), computed only over those 
lines detected individually with a significance $\ge 2\sigma$. 

%
\begin{table}
\footnotesize
\begin{center}
\caption{\bf \small Significance of the Absorption Systems}
\vspace{0.4truecm}
\begin{tabular}{|ccccc|}
\hline
Redshift & $\ge 2\sigma$ Ion & Probability of & Significance & Sum of $\ge 
2\sigma$ lines \\
& Detections & Exceeding $F$ & in $\sigma$ & significances (in $\sigma$) \\
\hline
$0.011 \pm 0.001$ & OVII, NVII & $5.9 \times 10^{-4}$ & 3.5 & 5.8 \\
$0.027 \pm 0.001$ & OVII, NVII, NVI & $1.3 \times 10^{-6}$ & 4.9 & 8.9 \\
\hline 
\end{tabular}
\end{center}
\end{table}
\normalsize
%

\section{The HST and FUSE Spectra of Mkn~421}

\subsection{HST-GHRS}
Mkn~421 was also observed with the Hubble Space Telescope (HST) GHRS on 
1995 February 1. 
An intervening HI Ly$\alpha$ system at $cz = (3046 \pm 2)$ km s$^{-1}$ 
was discovered and reported by Shull et al. (1996). We retrieved this GHRS 
spectrum of Mkn~421 from the public HST archive, and re-analyzed the data. 
We fitted the 1220-1250 \AA\ HST spectrum with a combination of broad 
continuum components and three narrow absorption gaussians to model the 
intervening HI Ly$\alpha$ and two strong Galactic Si lines visible in the 
spectrum (Figure 12, Table 2). 
Our best fitting parameters for the HI Ly$\alpha$ line (we assume the 
observed $b=40^{+5}_{-4}$ km s$^{-1}$ -- errors are at 3$\sigma$ -- 
to estimate the HI column) are 
consistent, within the uncertainties, with those reported by Penton et al. 
(2000a). 
The redshift of the HI Ly$\alpha$ line 
(Tab. 2) is consistent with that of the $z=0.011$ X-ray absorption system 
within the large systematic X-ray uncertainties ($\Delta z = 0.001$). 
We also computed the 3$\sigma$ HI column upper limit of a putative 
HI Ly$\alpha$ at the average redshifts of the two $z>0$ X-ray systems
(Tab. 2), and with $b(z=0.011) = 130$ km s$^{-1}$ and $b(z=0.027) = 185$ km 
s$^{-1}$ (i.e. $T(z=0.011) = 10^6$ and $T(z=0.027) = 2 \times 10^6$ K: see 
\S 7.4). 
We found N$_{HI}(z=0.011) < 4.7 \times 10^{12}$ cm$^{-2}$ and 
N$_{HI}(z=0.027) < 8.5 \times 10^{12}$ cm$^{-2}$ (Table 2). 
%
\begin{figure}
\hspace{-0.8in}
\epsfbox{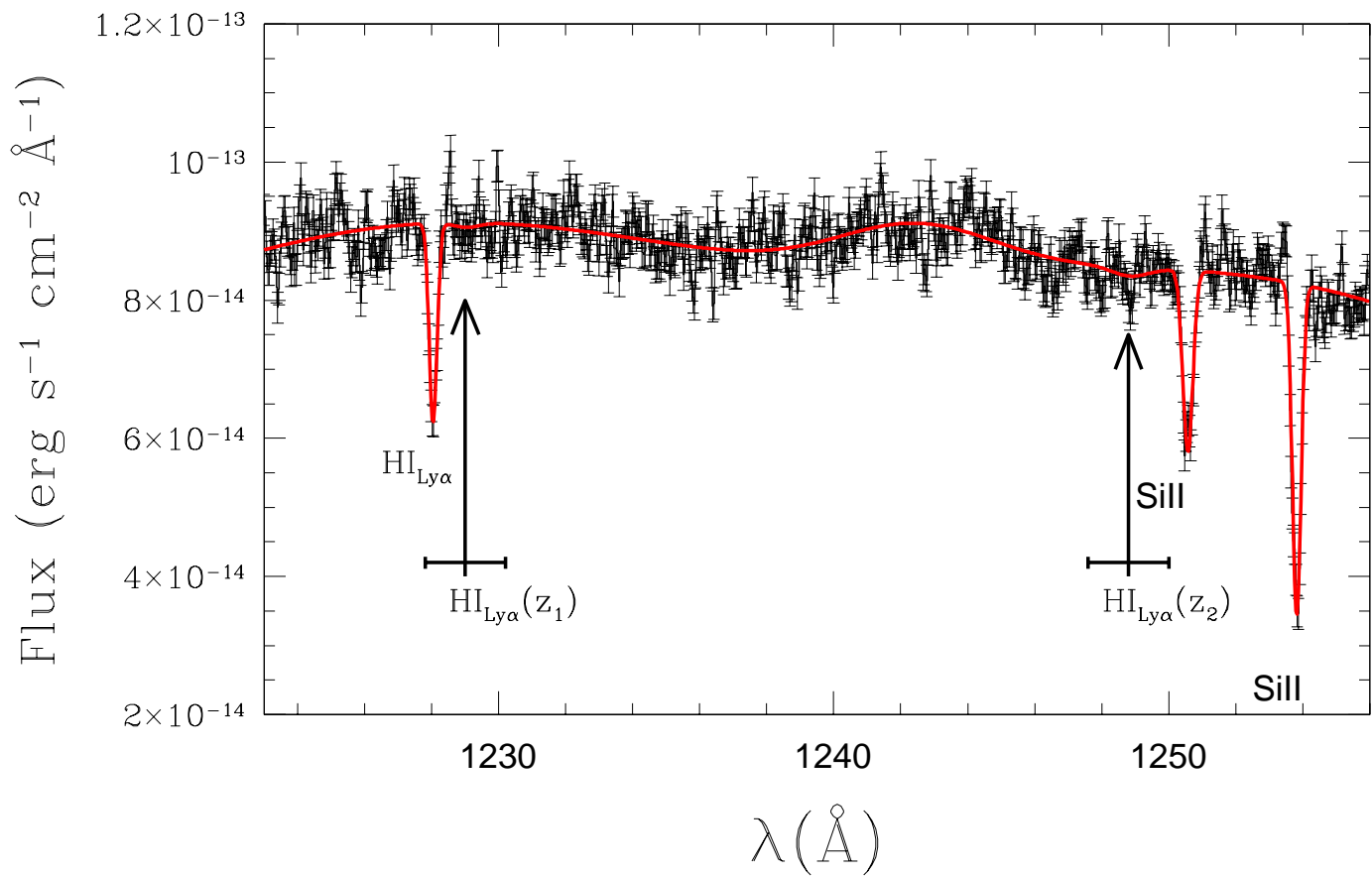}
\vspace{0in}\caption[h]{\footnotesize HST-GHRS spectrum of Mkn 421 between 
1223 and 1256 \AA. The best fitting continuum plus absorption line model is 
shown as a solid line. 
The two vertical arrows mark the positions of the HI Ly$\alpha$ 
at the redshifts of the two $z>0$ X-ray absorbers. The horizontal error bars 
on these arrow show the $\pm 300$ km s$^{-1}$ systematic uncertainty of the 
LETG.}
\end{figure}
%

\subsection{FUSE}
Following our TOO request Mkn~421 was observed by the Far Ultraviolet 
Spectrometer Explorer (FUSE) on 2003, January 19-21 (MJD = 52658-52660), 
with a total exposure time of 62 ks, based on a trigger from another period 
of strong X-ray activity as recorded by the RXTE-ASM (Figure 1). 

We searched the FUSE spectrum of Mkn~421 for OVI$_{2s\rightarrow2p}$
absorption at or close to the redshifts of our two $z>0$ X-ray absorbers
(Figure 13a,b). 
%
\begin{figure}
\hspace{-0.8in}
\epsfbox{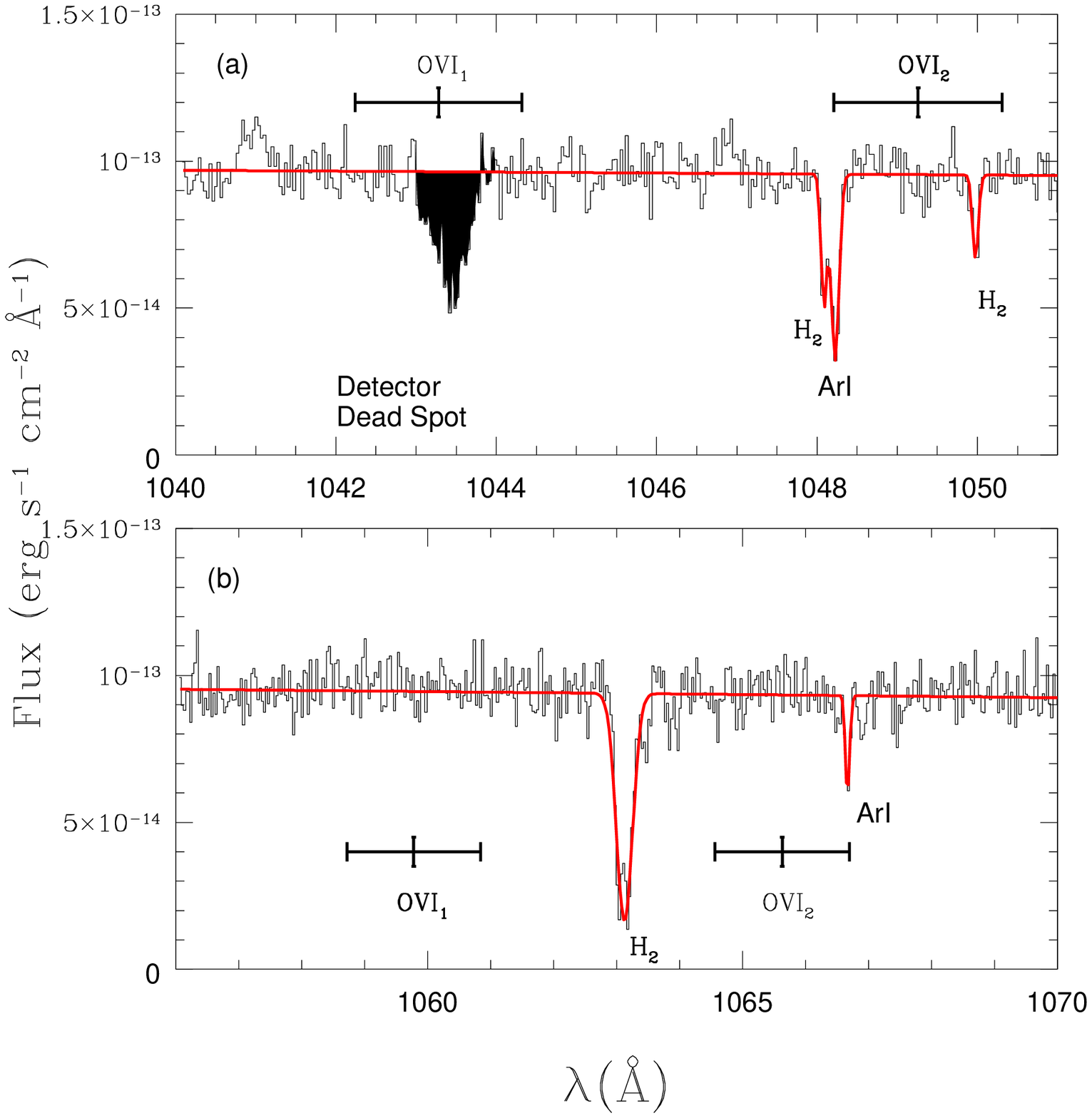}
\vspace{0in}\caption[h]{\footnotesize Two portions of the FUSE spectrum of 
Mkn 421 in the 1040-1051 \AA\ (upper panel) and 1056-1070 \AA\ 
(lower panel) ranges, along with their best fitting continuum plus 
absorption line models. 
The shaded area in Fig. 13a, indicates a detector flaw, due to a dead 
spot in the detector (Savage et al., 2005). 
The horizontal error bars in the two panels show the position of the 
expected lines of the OVI doublet at $z=0.011$ (Fig. 13a) and $z=0.027$ 
(Fig. 13b) $\pm 300$ km s$^{-1}$ systematic uncertainty of the LETG.}
\end{figure}
%
We fitted the two portions of the FUSE spectra between 1040 and 1051 \AA\ 
(Fig. 13a) and 1056 and 1070 \AA\ (Fig. 13b) with a combination of a 
polynomial continuum and a number of absorption gaussians to model 
the absorption features (mainly Galactic H$_2$ lines) visible 
in these two spectral regions. In particular, we fitted 2 H$_2$ lines 
and one ArI line between 1040 and 1051 \AA\ (Figure 13a), and one H$_2$ and 
one ArI line between 1056 and 1070 \AA\ (Figure 13b). 
We then added to our models two absorption gaussians with same FWHM, with 
their relative positions forced to the separation of the 
OVI$_{2s\rightarrow2p}$ 
doublet lines ($\lambda$(OVI$_1$)/$\lambda$(OVI$_2$) = 0.9945), and with 
their relative amplitudes forced to be 2, the relative amplitude of 
unsaturated OVI$_{2s\rightarrow2p}$ doublet lines (i.e. 
Ampl(OVI$_1$)/Ampl(OVI$_2$) = 2). 
For the X-ray system at z=0.011, the strongest line of the 
OVI$_{2s\rightarrow2p}$ doublet would fall at $\lambda = 1043.28$ \AA, 
where a detector artifact due to a dead spot (Savage et al., 2005; 
Figure 13a) hampers its detection. The position of the second line of 
the putative doublet would fall at $\lambda = (1049.03 \pm 0.05)$ 
(Figure 13a) where a moderate deficit in the continuum, compared 
to the best fit model, is visible. However, this region of the 
spectrum is potentially contaminated by Galactic H$_2$ absorption 
(4-0 R(0) at $\lambda = 1049.367$), so making the identification 
of a putative $\lambda=1037.617$ \AA\ OVI line at $z=(0.011 \pm 0.001)$ 
unreliable. 
Unfortunately, we cannot use the FUSE spectrum of Mkn 421 to 
measure OVI absorption at $z=(0.011 \pm 0.001)$ and therefore we only 
measure a 3$\sigma$ upper limit on the OVI column of N$_{OVI}(z=0.011) 
< 1.6 \times 10^{13}$ cm$^{-2}$ (Table 2). 
No OVI lines are visible at the redshift of the HI Ly$\alpha$ 
absorber, which gives a 3$\sigma$ upper limit on the column 
density of OVI at that redshift of N$_{OVI}(z=0.01) < 1.4 \times 10^{13}$ 
cm$^{-2}$ (Table 2). 
Similarly, we do not detect any OVI counterpart for the X-ray system at 
$z=0.027$, down to a 3$\sigma$ sensitivity of N$_{OVI} \le 1.4 \times 
10^{13}$ cm$^{-2}$ (Figure 13b, Table 2). 

\section{Summary of Absorption System Properties}

\subsection{The $z = 0.011$ Absorption System}
The $z=0.011$ X-ray system is mainly detected through OVII and NVII 
absorption. 
No NeIX, NVI or CVI absorption lines are detected at this redshift, while 
only a weak 3$\sigma$ upper limit can be set on OVIII absorption, due to 
the presence of a strong instrumental feature. Based on the detected lines 
we measure an average redshift of the system of $z = 0.011 \pm 0.001$, or 
$cz = (3300 \pm 300)$ km s$^{-1}$ (Table 1). Due to an instrumental flaw 
affecting the FUSE spectrum of Mkn 421 between $\lambda = 1043$ \AA\ and 
$\lambda = 1043.8$ \AA\ (Figure 13a), only a 3$\sigma$ upper limit can be 
set on the column of OVI at $z=0.011$. 
Finally, a strong HI Ly$\alpha$ 
is detected at a consistent redshift in the HST-GHRS spectrum of Mkn~421 
(Shull et al., 1996). 

\subsection{The $z = 0.027$ Absorption System}
The $z=0.027$ system has four identified detections of resonant 
absorption transitions from as many different ions (CVI, NVI, NVII and 
OVII), plus two uncertain EWs (because of the presence of two instrumental 
features) detections of NeIX and the OVII$_{K\beta}$, which we treat as 
$3\sigma$ upper limits. 
Based on the four detected lines we measure an average redshift of the 
system of $z = 0.027 \pm 0.001$. 
No HI Ly$\alpha$ or OVI$_{2s\rightarrow2p}$ absorption is detected in the 
GHRS and FUSE spectra of Mkn~421 at a redshifts of $z=0.027$. Assuming a 
temperature of $\sim 10^6$ K and pure thermal broadening of these lines 
(plus stretching due to the Hubble flow for an overdensity extending 2 Mpc 
along our line of sight), we estimated 3$\sigma$ upper limits of 47 m\AA\ 
and 18 m\AA\ respectively for these transitions, corresponding to 
N$_{HI} < 8.5 \times 10^{12}$ cm$^{-2}$ and N$_{OVI} < 1.4 \times 10^{13}$ 
cm$^{-2}$ (Table 2). 

\section{Local Galaxy Environment of the X-ray Absorbers}
To evaluate the density of the galaxy environment in which 
the X-ray absorbers lie, we searched a number of galaxy catalogs sensitive 
down to $m_b = 15.5$: the CfA Redshift Catalog (v.June 1995, CfA95
\footnote{http://cdsweb.u-strasbg.fr/viz-bin/Cat?VII/193}
), the CfA-North-Galactic-Pole+36 Catalog (CfA-NGP+36, 
Huchra et al., 1995, ApJS, 99, 391) and the Updated Zwicky Catalog (UZC99, 
Falco et al., 1999, PASP, 111, 438). The CfA-NGP+36 completes, 
in redshift identifications, the CfA95 catalog in the declination range 
32.5$^0$-38.5$^0$ and right ascension range $8^h < \alpha < 17^h$ 
(which includes the position of Mkn~421), and down to $m_b = 15.5$. 
The UZC99 catalog corrects some of the inadequacies of the previous 
CfA catalogs, which were all based on the Zwicky Catalog of galaxies 
and relied on heterogeneous set of galaxy positions and redshifts. 

For the $z=0.011$ system, we found 4 galaxies (2 of which forming a close 
system, in interaction) in a comoving cylinder with a base-radius of 5 Mpc, 
and half-depth of 500 km/s (in the $z$ dimension: $2800 \le cz \le 3800$ km 
s$^{-1}$; Figure 14; see also Penton et al., 2000b)
\footnote{This depth correspond to a physical size of the 
cylinder in the $z$ direction, of $\sim 14$ Mpc, if peculiar galaxy motions 
are neglected.} 
\footnote{The linear sizes of this volume, correspond to the typical 
sizes expected for the predicted intergalactic filaments connecting virialized 
structures in the local Universe (i.e. Cen \& Ostriker, 1990, Hellsten et al., 1998).}
. All these galaxies have $\Delta (cz) < 140$ km s$^{-1}$ from the $z=0.011$ 
system (within the absorber redshift uncertainty), and lie between 4.4 and 
5.1 Mpc from it. 
At the absorber distance, the limit magnitude of the catalogs, translates 
into an absolute magnitude limit of M$_b = -17.9$, i.e. 2.75 mag fainter than 
M$^*_B(SDSS)$ (Poli et al., 2003
\footnote{Poli et al. (2003) derive M$^*_B(SDSS)= -20.65$ from the g magnitude 
local Luminosity Function of galaxies from the Sloan Digital Sky Survey (SDSS, 
Blanton et 
al., 2001) adopting the following transformation M$^*_B = g* + 0.32$ and scaling 
magnitude and normalization to H$_0 = 70$ km s$^{-1}$ Mpc$^{-1}$.}
). This implies a density of galaxies at $z=0.011$, down to $M_b = -17.9$, of 
$\phi_{obs}^{z=0.011}(M_b\ge -17.9) = 3.6 \times 10^{-3}$. 
We used the transformed $g^*$ to B band SDSS local Luminosity Function of 
galaxies (see Fig. 1 in Poli et al., 2003), to compute the local density 
of galaxies down to this magnitude limit, and found $\phi_{SDSS}(M_b\le 
-17.9) = 0.017$ Mpc$^{-3}$. This gives a relative density of galaxies in our 
region of Universe centered on the $z=0.011$ system, of 
$\phi_{obs}^{z=0.011}(M_b\ge -17.9)/ \phi_{SDSS}(M_b\le -17.9) = 0.2$. 
The $g^*$ band SDSS luminosity function is an average over 4684 galaxies 
in a volume $\sim 5.1 \times 10^6 h_{0.7}^{-3}$ Mpc$^3$ (Blanton et al., 2001) 
and so is a good mean sample of the local Universe. We conclude that this system 
lies in an under dense volume of Universe. 
%
\begin{figure}
\hspace{-0.8in}
\epsfbox{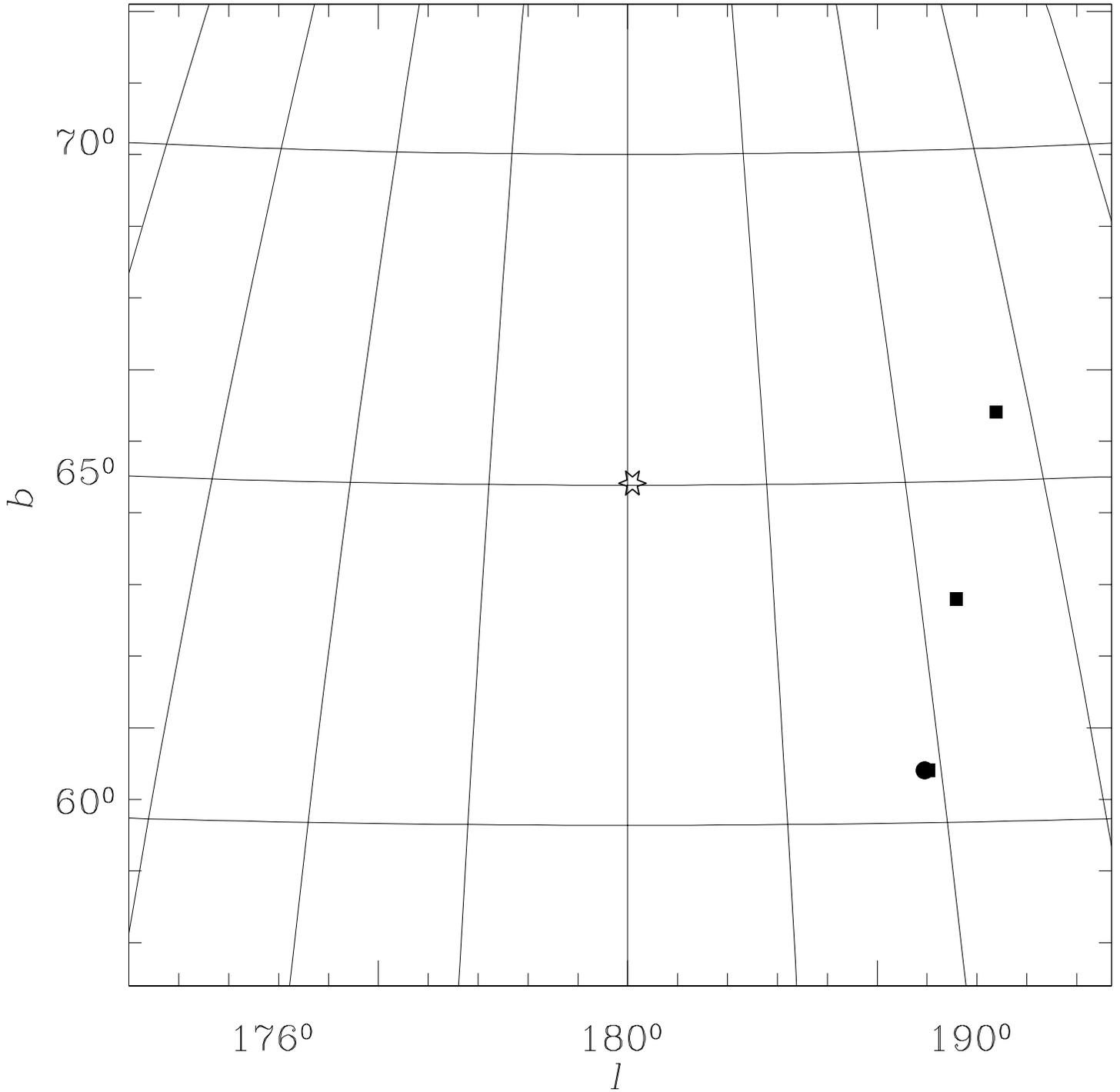}
\vspace{0in}\caption[h]{\footnotesize Aitoff plot of a region of sky, in 
Galactic coordinate around the $z=0.011$ X-ray 
absorber. Four galaxies are found in a cylinder centered on the 
absorber, with a base radius of 5 Mpc and half-depth of $\Delta(cz) = 500$ km 
s$^{-1}$ ($\sim 7$ Mpc), down to a magnitude limit of 
$m_b = 15.5$. Squares and circles indicate blueshifted and redshifted 
galaxies respectively, compared to the absorber redshift. 
The star marks the position of the absorber.}
\end{figure}

\medskip
The $z=0.027$ system lies $\sim 13$ Mpc from Mkn~421. 
The CfA95, CfA-NGP+36 and UZC99 catalogs contain a total of 8 different 
galaxies down to $m_b = 15.5$ in a comoving cylinder with a base-radius of 
5 Mpc and half-depth of 500 km/s (in the $z$ dimension: $7600 \le cz \le 
8600$; Figure 15)$^{12\hbox{ }13}$. 
These galaxies lie between 3.3 and 8.2 Mpc from the absorber. 
At the absorber distance, the limiting magnitude of the catalogs, translates 
into an absolute magnitude limit of M$_b = -19.8$, or 0.85 mag fainter than 
M$^*_B(SDSS)$ (Poli et al., 2003). The observed density of 
galaxies at $z=0.027$, down to this magnitude limit, is then 
$\phi_{obs}^{z=0.027}(M_b\le -19.8) = 7.3 \times 10^{-3}$ Mpc$^{-3}$. 
The local density of galaxies for the same magnitude limits, from the 
B band SDSS local luminosity function of galaxies (Poli et al., 2003), is also 
$\phi_{SDSS}(M_b\le -19.8) = 7.3 \times 10^{-3}$ Mpc$^{-3}$, which implies a 
relative density of galaxies of $\phi_{obs}^{z=0.027}(M_b\le -19.8)/
\phi_{SDSS}(M_b\le -19.8) = 1$. The $z=0.027$ system seems to lie in an 
average density region. 
%
\begin{figure}
\hspace{-0.8in}
\epsfbox{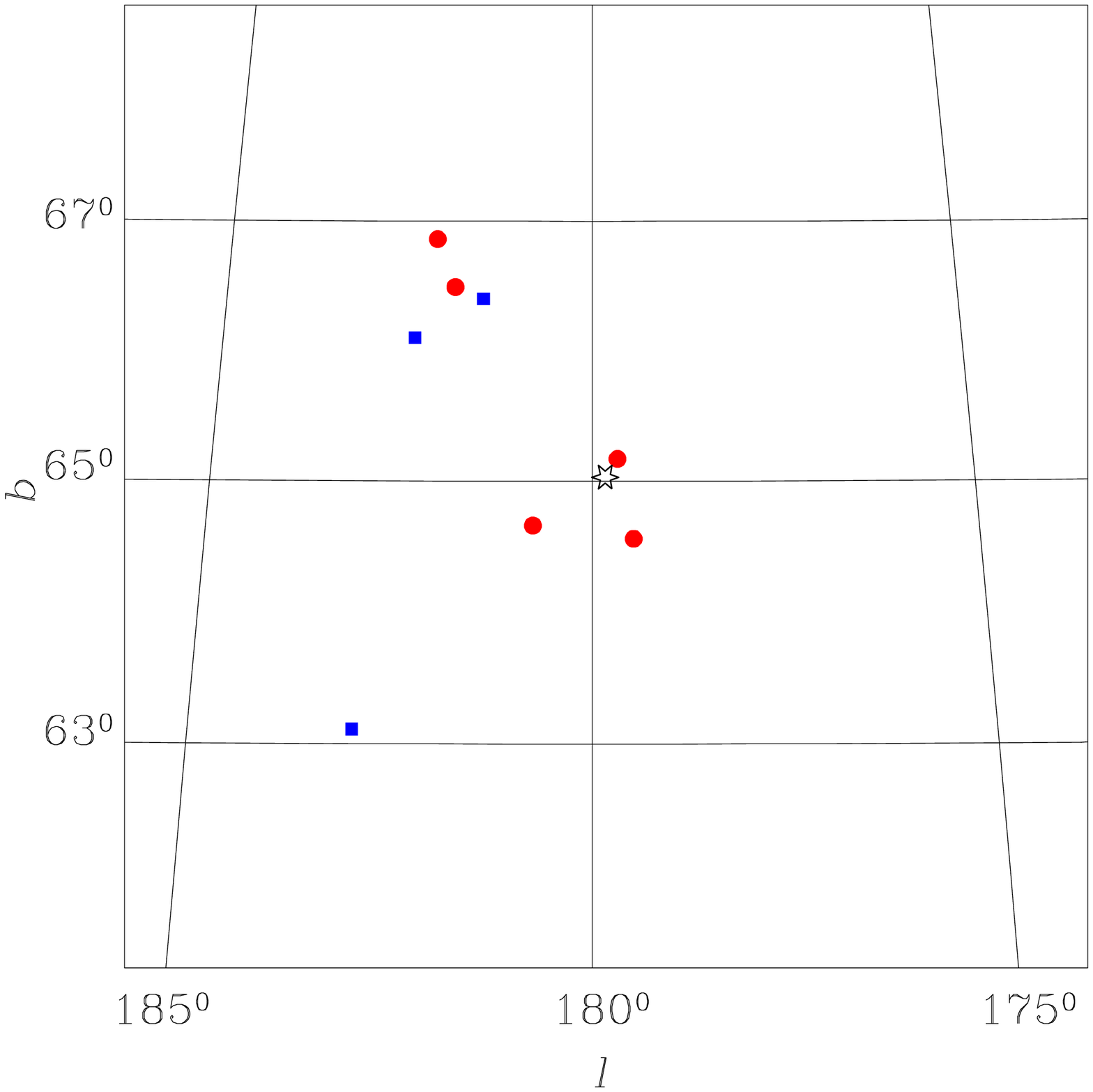}
\vspace{0in}\caption[h]{\footnotesize Aitoff plot of a region of sky, in Galactic 
coordinate around the X-ray absorber at $z = 0.027$. 
Eight galaxies are visible in a cylinder centered on the absorber, with a 
base radius of 5 Mpc and half-depth of $\Delta(cz) = 500$ km s$^{-1}$ ($\sim 7$ Mpc), 
down to $m_b = 15.5$. Squares and circles indicate 
blueshifted and redshifted galaxies respectively, compared to the absorber 
redshift. The star marks the position of the absorber.}
\end{figure}
%

\section{Discussion}

\subsection{Line Saturation and EW Ratio Diagnostics}
In neither X-ray absorption system do we detect more than one line from the same ion. 
So we can not use the standard EW ratios technique to estimate line 
saturation directly. 
We instead evaluated line-saturation by using the standard alternative 
Curve-of-Growth (CoG) technique. We allowed the Doppler 
parameter to vary between $b = 50$ and $b = 400$ km s$^{-1}$. Since all the 
lines are unresolved we can place a 2$\sigma$ limit of $b < 480$ km s$^{-1}$. 
The largest 3$\sigma$ determinations of our line EWs, 
for both systems, fell well within the linear branch of the corresponding 
CoGs even for $b = 50$ km s$^{-1}$, and far from saturation. 

On the linear branch the ratio of the EWs of two lines from two ions 
$X^i$ and $Y^j$ of the two elements $X$ and $Y$, is directly proportional 
to the ratio between the corresponding relative ion abundances 
$\xi_{X^i}/\xi_{Y^j}$: 
\begin{equation}\label{ionew}
{\xi_{X^i} \over \xi_{Y^j}} = {W_{X^i}(m\AA) \over W_{Y^j}(m\AA)} 
\left({\lambda(Y^j) \over \lambda(X^i)}\right)^2 {f_{Y^j} \over f_{X^i}} 
{A_Y \over A_X}. 
\end{equation}. 
In eq. \ref{ionew} $f_{Y^j}/f_{X^i}$ is the oscillator strength ratio 
for the given transition, and $A_Y/A_X$ is the relative abundance of 
element $Y$ compared to $X$. 
Two or more of these ratios from ions of the same element, then, uniquely 
define the ionization balance of the absorber, and so allow us to derive 
the physical conditions in the gas, as well as the ionization mechanism 
at work. 
The ratio of the EWs of lines from ions of different elements also depends 
on the ion abundance and so constrains the chemical composition of the gas. 

We used Cloudy (vs 90.04, Ferland et al., 1997) to build grids of hybrid, 
collisional ionization plus photoionization models, as a function 
of the electron temperature, to compare predicted ion relative abundance 
ratios with observed EW ratios. 
These models reduce to pure collisional models for electron 
density of the gas $n_e >> 10^{-6}-10^{-5}$ cm$^{-3}$.

\subsection{Modeling the Photoionization Contribution}
Following Nicastro et al. (2002), we used the diffuse UV-to-X-ray background 
(XRB, hereinafter) as the photoionizing source. 
We also included the possible 'proximity effect' (c.f. HI Ly$\alpha$, 
Bechtold, 1994) of photoionization by the 
blazar Mkn~421. A 100 mCrab flux for Mkn~421 implies beamed X-ray luminosities 
along our line of sight, of the order of $L_{Mkn~421} \sim 10^{45}$ erg 
s$^{-1}$. Such luminosity can photoionize low density gas along our line of sight 
to Mpc-scale distances. Assuming the observed UV-to-X-ray Spectral 
Energy Distribution (SED) of Mkn~421, the ionization parameters
\footnote{$U$ is here defined as the ratio between the radiation field 
ionizing photon density at the distance of the gas cloud, and the electron 
density of the gas, and is computed assuming constant density through  
the cloud, and a plane parallel geometry (i.e. thickness of 
the cloud along the line of sight, much smaller than the distance of the 
cloud from the ionizing source).}
, at the illuminated face of the clouds, are $U_{Mkn~421} 
\simeq 1.7 \times 10^{-2} \delta^{-1}$ at $z=0.011$ and $U_{Mkn~421} 
\simeq 0.8 \delta^{-1}$ at $z=0.027$ (we have used $\Omega_b = 0.04$ and $h = 0.7$). 
The recombination time for OVII in photoionized gas with out-of-equilibrium 
temperature of $T=10^6$ K and overdensity $\delta \ls 500$ is $t_{rec} 
\ge 2.4 \times 10^8 \delta_{500}^{-1}$ years (e.g. Nicastro et al., 1999, 
and references therein), and increases/decreases for lighter/heavier 
elements. 
The inverse, equilibrium 'photoionization' time is typically more than one 
order of magnitude shorter, and depends on the variability timescale of the 
ionizing source. 
Given the frequency of the $100$-mCrab type bursts recorded from 
Mkn~421 ($\sim 1.5$/year in the past 7 years), and their typical duration 
of $\sim 3$ weeks, we must conclude that any intervening IGM condensation 
within $\sim 10-20$ Mpc of this blazar with overdensities $\delta \ls 500$ 
must be kept in a relatively high ionization state due to this proximity 
effect. Moreover, given the different times with which different ions 
react to ionizing flux changes the overall ionization balance in such a gas 
could be significantly different from the one expected at equilibrium. 
The photoionization contribution due to the diffuse XRB 
can be parameterized, for both absorbers, by the ionization parameter 
$U_{XRB} \sim 0.15 \times \delta^{-1}$ (again, assuming $(\Omega_b h^2/0.02)
 = 1$, and neglecting the small redshift difference between the two 
systems). 
By comparing the proximity effect with the ionization induced 
by the UV-X-ray background for the two systems we conclude that 
photoionization by Mkn~421 is completely negligible for the filament at 
$z = 0.011$ , while it dominates the overall photoionization contribution 
by a factor of $\sim 5$ for the filament at $z = 0.027$.  

\subsection{Self-Consistent N$_H$ versus T Solutions}
For both the $z=0.011$ and $z=0.027$ absorption systems we compared the 
measurements (X-rays) or 3$\sigma$ upper limits (UV, FUV and X-rays) on the 
EW ratios (\S 7.1), with the predicted ion relative abundance 
ratios from our grid of hybrid collisional-ionization plus photoionization 
models (\S 7.2), to constrain maximal intervals of temperatures defining 
common solutions (e.g. Nicastro et al., 2002). 
The existence of a common range of temperatures for all ion ratios, however, 
still does not guarantee that a range of self-consistent equivalent 
hydrogen column density versus temperature solutions exist. 
To search for these solutions we used the each measured (or 3$\sigma$ upper 
limit) equivalent width $W_X^i$ for a given ionic transitions to estimate 
a temperature (and metallicity) dependent equivalent hydrogen column 
density N$_H(T,[X/H])$ (where $[X/H]$ is the logarithm of the metallicity 
of the generic element $X$, compared to solar). The temperature dependence 
is introduced by the dependence of N$_H$ upon the relative ion abundance 
$\xi_i$, in the N$_H$ versus $W_X^i$ relationship (see e.g. eq. (1)). 
This way we were able to define two ranges of self-consistent N$_H$ versus 
T solutions for the two systems, and constrain their metallicity 
(\S 7.4 and 7.5). 

\subsection{Physical State of the Absorber at $z = 0.011$}
The only X-ray absorption line of the $z=0.011$ system individually 
detected at a significance larger than 
3$\sigma$ is the OVII$_{K\alpha}$ line at 21.85 \AA\ (Table 2). So we 
use this ion as the reference ion for the ionization balance 
diagnostics. 

The predicted ratios for several ions compared to OVII, for hybrid 
collisional-ionization plus photoionization gas, are shown as thin solid 
curves in Figure 16, plotted against the gas temperature: 
(a) OVI/OVII (black curve), 
(b) OVIII/OVII (red curve), (c) NVII/OVII (blue curve), and (d) HI/OVII 
(green curve). The predicted ratios are for a baryon density of 
$n_b = 10^{-5}$ cm$^{-3}$ ($\delta = 50$). 
The thick segments superimposed on these curves are our $\pm 3\sigma$ 
allowed ranges for the corresponding ratios. 
We assume [O/H] = -1 and [N/O] = 0. Lower values of [O/H] would 
further reduce the allowed range of temperatures set by the HI/OVII 
interval (green segment on the green curve), while larger [O/H] would broaden 
the interval. 

First we note that the observed width of the HI Ly$\alpha$ at $\lambda = 
1228.02 \pm 0.01$ (Figure 12, Table 2) sets a 3$\sigma$ upper limit on the 
temperature of the HI absorber (dashed vertical line in Figure 16) of 
$T_{HI} = 1.2 \times 10^5$ K. 
This is incompatible with the, metallicity independent, lower limit of 
$T_{FUV-X-ray} = 5.3 \times 
10^5$ K on the temperature of the FUV-X-ray absorber, imposed by the 
measured 3$\sigma$ lower boundary on the OVI/OVII ratio (Fig. 16, left end 
of the OVI/OVII black segment). 
Decreasing the overdensity of the FUV-X-ray absorber from the adopted value 
of $\delta=50$ to, say, $\delta = 10$, would slightly flatten the OVI/OVII 
curve at low temperatures, extending the OVI/OVII lower end interval to 
$T_{FUV-X-ray} \sim 3.9 \times 10^5$ K. This reduces the 
inconsistency between the temperatures of the HI and FUV-X-ray absorbers, 
but they remain mutually incompatible. We conclude that if the HI and 
FUV-X-ray systems are related, they must co-exist in a multiphase IGM. 
In the following we use the HI Ly$\alpha$ 3$\sigma$ upper 
limit at $\lambda = 1229.04$ \AA\ ($z=0.011$, Table 2) as a constraint for 
condition in the FUV-X-ray absorbing gas. 

Analogously to the OVI/OVII ratio (which sets a minimal temperature of 
$T_{FUV-X-ray}^{min} = 5.3 \times 10^5$ K), the measured OVIII/OVII 
3$\sigma$ upper limit sets a metallicity independent maximal temperature of 
the absorber at $T_{FUV-X-ray}^{max} \le 3.3 \times 10^6$ K (Fig. 16, right 
end of the OVIII/OVII red segment). 

On the contrary the measured (3$\sigma$) HI/OVII and NVII/OVII ratios 
depend upon metallicity. In Figure 16 the thick green and blue segments 
refer to Solar and $0.1 \times$ Solar metallicity respectively. These 
measurements define broad ranges of temperatures that both intersect the 
metallicity independent maximal range defined by the OVI/OVII and 
OVIII/OVII ratios. 

Additionally, the 3$\sigma$ upper limit on the NVII/OVIII ratio constrains 
the [O/N] relative abundance to be [O/N]$\ge -2.05$, consistent with solar
\footnote{From eq. \ref{ionew}, the relative metallicity between two elements 
$Y$ and $X$ depends on the inverse relative ion abundance ratio 
$\xi_{X^i}/\xi_{Y^j}$, and on the direct EW ratio $W_{Y^j}/W_{X^i}$. 
For a $W_{X^i}$ $3\sigma$ upper limit, this relative metallicity has to be 
always larger than the 3$\sigma$ lower limit on the ratio $W_{Y^j}/W_{X^i}$, 
multiplied by the minimum allowed value of the curve $\xi_{X^i}/\xi_{Y^j}$.}
. Similarly, and more importantly, we can derive a lower limit on the 
oxygen metallicity ratio. The non-detection of HI Ly$\alpha$ together 
with the detection of OVII K$\alpha$ and the minimum amount of relative HI 
to OVII abundance predicted by the model, set a stringent lower limit on 
the oxygen to hydrogen metallicity ratio, of [O/H]$\ge -1.46$. 

%
\begin{figure}
\hspace{-0.8in}
\epsfbox{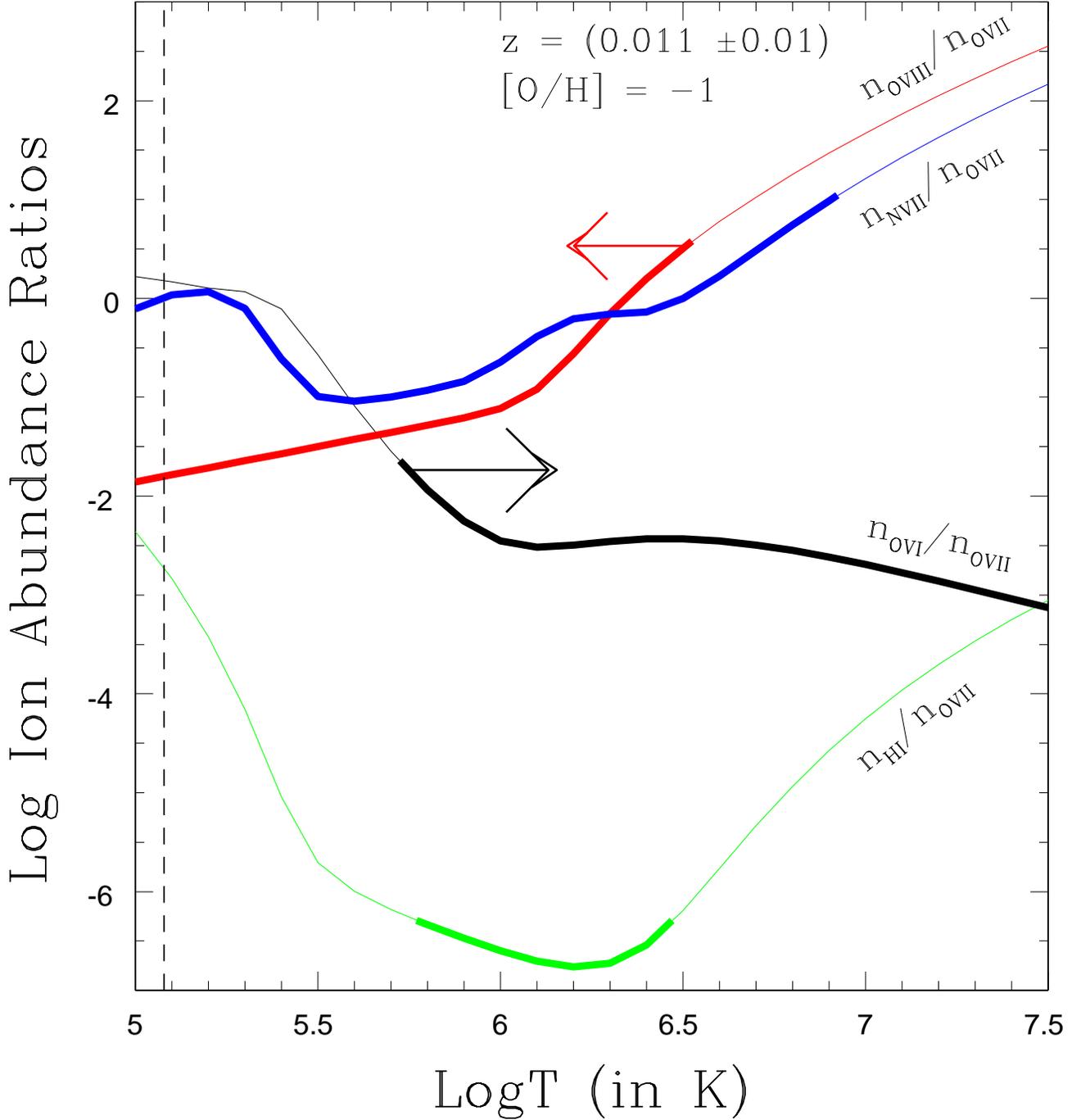}
\vspace{0in}\caption[h]{\footnotesize Predicted ratios for several ions 
compared to OVII, for hybrid collisional-ionization plus photoionization 
gas (thin solid curves), plotted against the gas temperature: (a) OVI/OVII 
(black curve), (b) OVIII/OVII (red curve), (c) NVII/OVII (blue curve), and 
(d) HI/OVII (green curve). 
The predicted ratios are for a baryon density of $n_b = 10^{-5}$ cm$^{-3}$ 
($\delta = 50$). 
The thick segments superimposed on these curves are our $\pm 3\sigma$ 
allowed ranges for the corresponding ratios for the $z=0.011$ system. 
We assume [O/H] = -1 and [N/O] = 0.
The dashed vertical line sets a 3$\sigma$ upper limit on the 
temperature of the HI absorber, from the width of the HI Ly$\alpha$ at 
$\lambda = 1228.02 \pm 0.01$ (Figure 12, Table 2).}
\end{figure}
%

The two arrows in Figure 16 shows the extremes of the maximal range of 
temperatures of the $z=0.011$ FUV-X-ray absorber: $T_{FUV-X-ray} \in 
[0.53-3.3] \times 10^6$ K. 
Coupling this maximal temperature range with the estimates on 
N$_H$ obtained from each available ion (see \S 7.3), 
we find that a common range of N$_H$ versus 
$T_{FUV-X-ray}$ solutions exists in the temperature interval 
$T_6 \in [0.6-2.5]$, and is given by: N$_H(z=0.011) = 
[(1.7 \pm 0.7)T_6 - (0.1 \pm 0.6)] \times 10^{19} \times 10^{-[O/H]_{-1}}$ 
cm$^{-2}$ (where $T_6$ is the temperature of the absorber in units of 
$10^6$ K). 
This implies a thickness of the absorber along our line of sight 
$D(z=0.011) = $N$_H / n_b = [(0.55 \pm 0.23)T_6 - (0.03 \pm 0.19)] 
(n_{-5b})^{-1} 10^{-[O/H]_{-1}}$ Mpc, assuming constant baryon density. 
In this equation $(n_{-5b})$ is the baryon density in units of $10^{-5}$ 
cm$^{-3}$. 

We notice here that due to the weakness of the OVIII/OVII and NVII/OVII 
ratio constraints, our ionization balance solutions are rather insensitive 
to even factor 10 differences in the baryon density. 
Higher signal to noise X-ray data are needed to constrain 
both the OVIII and the NVII columns to better than 30-40 \% at 2$\sigma$, 
and so set limits on the value of the baryon density. This is obtainable 
with 2.5-3 times the number of counts we currently have at the wavelengths 
of the OVIII and NVII K$\alpha$ transitions. 

\subsection{Physical State of the Absorber at $z = 0.027$}
Two X-rays lines of the $z=0.027$ system are detected at a conservative 
significance $\ge 3\sigma$: the NVII$_{K\alpha}$ at $\lambda = 25.44$ \AA\ 
and the NVI$_{K\alpha}$ at $\lambda = 29.54$ \AA\ (Table 1). In our 
ionization balance diagnostics analysis we use NVII as the reference ion 
as it has slightly higher significance. 

As for the $z=0.011$ system we explore the hybrid collisional ionization 
plus photoionization case for the $z=0.027$ system (Figure 17). 
Seven EWs ratio constraints from 8 different ions are now available from the 
UV-FUV-X-ray determinations or 3$\sigma$ upper limits (Figure 17). 

First we note that the NVI/NVII constraint (blue interval in Figure 17) sets 
a tight and, most importantly, metallicity independent lower limit on the 
temperature of the absorber, of logT$\ge 5.98$ (vertical black solid line 
and blue arrow in Fig. 17). 
Based on the simple equilibrium models we adopt, however, a maximal 
temperature range of solutions can be 
found only for rather large values of the N/O ratio, compared to Solar. 
Solid intervals on the O/NVII curves in Fig. 17 are for [N/O] = 0, while 
broader dashed intervals are for [N/O] = 0.65. This is the minimum N/O 
overabundance, compared to the solar value, that reconciles the large 
observed absorption by He-like and H-like N with the lack, or weakness, 
of OVI, OVII and OVIII absorption. 
In particular, from the high temperature side (logT$\gs 6.4$ K), 
[N/O]$\ge 0.65$ allows the measured OVI/NVII and OVII/NVII upper limit 
ratios to exceed the predicted OVII/NVII local maxima at the knee 
temperature of $T\sim 10^{6.4}$ K, where the curve changes its slope and 
becomes much flatter. On the other hand, from the low-temperatures side 
(logT$< 6$ K), only for [N/O]$\gs 0.65$ can the OVIII/NVII constraint 
extend up to the logT$> 6$ K region, and so become compatible 
with both the OVI/NVII and the OVII/NVII constraints. 
A large N abundance is also consistent with the somewhat narrow range of 
allowed temperature that is derived from the HI/NVII measurement. 
The red interval 
on the theoretical HI/NVII curve is for [N/H] = -0.35, which together 
with [N/O] = 0.65, implies [O/H] = -1. At this relatively large (but still 
sub-Solar) [N/H] metallicity ratio, the temperature of the system is 
constrained between $5.97 <$ logT $< 6.49$ K. Larger or lower [N/H] values 
would broaden or narrow this interval. The minimum N/H metallicity 
ratio for this system is [N/H]$\ge -0.67$ (i.e. [O/H]$\ge -1.32$, for 
[N/O] = 0.65) at logT = 6.2 K. Finally the N/C ratio is consistent with 
the Solar value while only a moderate N/Ne overabundance of [N/Ne] = 0.08 
is required (so maybe indicating that also C and Ne are particularly 
abundant in this absorbed, compared to H). 

We stress that the above limits on the metallicity ratios depend critically 
on modeling, and so on (a) the density of the absorber, and (b) whether 
or not equilibrium photoionization (acting on gas already collisionally 
ionized through shocks) applies, and applies to all elements (we recall that 
recombination times are different for different elements). 
The predicted relative abundance ion ratios plotted in Fig. 17 are for a 
baryon density of $n_b = 10^{-5}$ cm$^{-3}$ ($\delta = 47$). Larger 
densities would generally increase all predicted ratios, so making the 
N overabundance problem even more serious. Lower densities, however, 
would generally lower the OVI/NVII and OVII/NVII knees at logT$\sim 6.4$ K, 
but would increase the OVIII/NVII ratio at low temperature: at $n_b = 
2 \times 10^{-6}$ cm$^{-3}$ ($\delta \sim 10$), a [N/O] = 0 solution can 
be found for all ions, except OVIII, with a temperature 
of logT$\sim 6.2$ K. Finally, the gas could be far from 
equilibrium, in which case no conclusion could be drawn of course 
on metallicity and temperature of the absorber based on our comparison 
with hybrid equilibrium models. 
%
\begin{figure}
\hspace{-0.8in}
\epsfbox{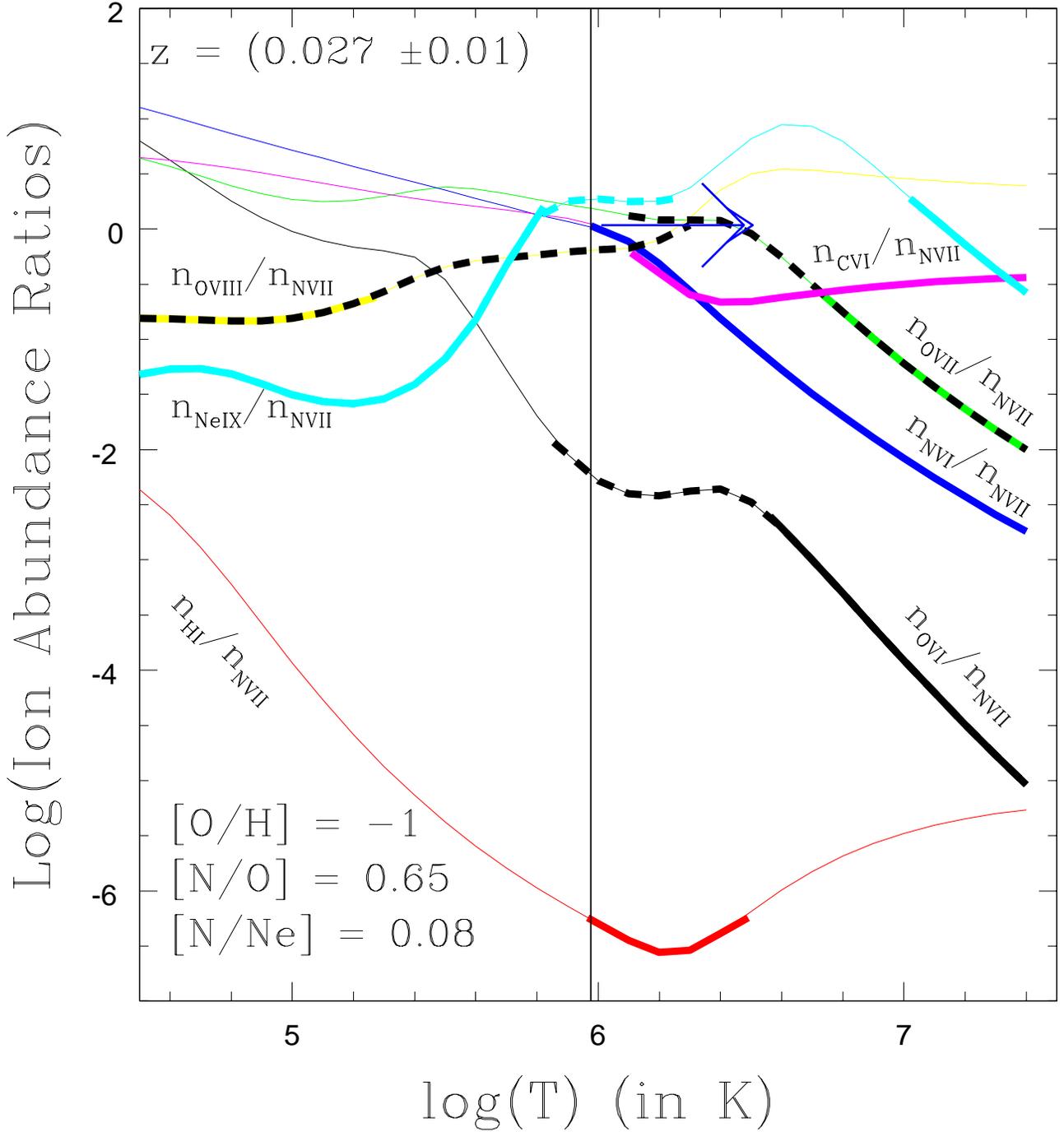}
\vspace{0in}\caption[h]{\footnotesize Same as Fig. 16, for the $z=0.027$ 
system. 
Predicted ratios for 7 different ions are compared NVII, for a baryon 
density of $n_b = 10^{-5}$ cm$^{-3}$ ($\delta = 50$). 
We assume [O/H] = -1 and [N/O] = 0 (solid interval) and 0.65 (dashed 
intervals). 
The NVI/NVII constraint (blue interval) sets 
a tight and metallicity independent lower limit on the 
temperature of the absorber, of logT$\ge 5.98$ (vertical black solid line 
and blue arrow).}
\end{figure}
%

For $[N/H] = -0.35$ and $[N/O] \ge 0.65$ the maximal temperature range 
for which N$_H$ versus $T$ are in principle allowed is set, on the lower 
side by the metallicity independent NVI/NVII ratio, and on the upper side 
by the metallicity dependent HI/NVII ratio: $T_6 \in [0.95-3.1]$. 
Coupling this maximal temperature range with the estimates on 
N$_H$ obtained from each available ion (see \S 7.3), 
we find that, for [N/O] = 0.65, a common range of N$_H$ versus 
$T$ solutions exists in the narrow (but, again, metallicity dependent) 
temperature interval $T_6 \in [1.2-1.6]$, and is given by: N$_H(z=0.027) = 
[(1.67 \pm 0.76)T_6 - (0.71 \mp 1.15)] \times 10^{19} \times 
10^{-\{[O/H]_{-1} + [N/O]_{0.65}\}}$ cm$^{-2}$. 
This implies a thickness of the absorber along our line of sight 
$D(z=0.027) = $N$_H / n_b = [(0.53 \pm 0.24)T_6 + (0.23 \mp 0.37)] 
(n_{-5b})^{-1} 10^{-\{[O/H]_{-1} + [N/O]_{0.65}\}}$ Mpc, assuming constant 
baryon density. 

\subsection{Two WHIM Filaments at $z=0.011$ and $z=0.027$}
The $z=0.011$ and $z=0.027$ absorbers that we detect 
along the line of sight to Mkn~421 have temperature ranges in good agreement 
with the theoretical prediction for the WHIM. 
The WHIM temperature distribution peaks at $\sim 4 \times 10^6$ K 
and is highly skewed toward lower temperature, with $T(\sigma_{-})/T_{peak} 
\sim 27$ and $T(\sigma_{+})/T_{peak} \sim 5$ (Dav\'e et al., 2001: see their 
Fig. 5). 
The limits that we set on the equivalent H columns of these two absorbers 
(N$_H(z=0.011) = [(1.7 \pm 0.7)T_6 - (0.1 \pm 0.6)] \times 10^{19} \times 
10^{-[O/H]_{-1}}$ cm$^{-2}$ and N$_H(z=0.027) = [(1.67 \pm 0.76)T_6 - (0.71 
\mp 1.15)] \times 10^{19} \times 10^{-\{[O/H]_{-1} + [N/O]_{0.65}\}}$ 
cm$^{-2}$) are also in excellent agreement with WHIM 
predictions (e.g. Hellsten et al., 1998, Dav\'e et al., 2001). 
Predictions for WHIM metallicities are still 
uncertain, mostly due to the lack of, or partial, non self-consistent, 
treatment of galaxy-IGM feedback mechanisms in current hydrodynamical 
simulations for the formation of large scale structure. However, values of 
about [O/H] = -1 or even larger are plausible at very low redshift, 
consistent with the $[O/H] \sim -1$ we have found. 
Moreover, the number of detected OVII systems is consistent, within the 
1$\sigma$ errors, with the predicted number of OVII WHIM filaments along 
a random line of sight, down to the lowest detected OVII column of 
N$_{OVII} = 7 \times 10^{14}$ cm$^{-2}$, and up to the distance of Mkn~421 
(see next section). 

We therefore identify the two absorption systems at $z=0.011$ and $z=0.027$ 
with two intervening WHIM filaments, with roughly same column and 
extent along our line sight. These are the first $\ge 3.5\sigma$ 
(conservative, or $\ge 5.8\sigma$ - sum of lines significance) detections 
of WHIM filaments in the X-rays. 

\subsubsection{Ruling Out Blazar Outflows}
We should also consider the possibility that the two absorbers are due 
to highly ionized material outflowing from either the nucleus of Mkn~421 
or the blazar's host galaxy at velocities of $\sim 5600$ km s$^{-1}$ and 
$\sim 900$ km s$^{-1}$ respectively. Even the lower of these two velocities 
is extreme for ISM clouds, when compared with the typical range of velocities 
observed in High Velocity Clouds (either cold, e.g. Blitz et al., 
1999, or warm-hot, e.g. Sembach et al., 2003) in our Galaxy. 
Few hundreds to 1000-2000 km s$^{-1}$ photoionized outflows are, instead, 
common in low redshift Seyfert 1s (e.g. Kriss, 2002), but have not been 
seen in blazars. 
Early reports of X-ray absorbers in blazars (Kruper \& Canizares, 1982; 
Madejski et al. 1991) have not been confirmed by later, higher spectral 
resolution, observations (e.g. Nicastro et al., 2002, Fang et al., 2003, 
Cagnoni et al., 2004). Moreover all photoionized outflows in Seyfert 1s 
are detected in absorption both in the UV (through CIV and OVI) and 
X-rays (through hundreds of transitions from ionized metals, e.g. Krongold 
et al., 2003, and references therein), while in neither of our two 
cases we do detect associated OVI. 

In the following we investigate this possibility further. If the absorber 
was part of the blazar environment, assuming a range of density of 
$10^6-10^{11}$ cm$^{-3}$ (typical of different gaseous components in an 
AGN environment, from the parsec-scale dusty torus, to the sub-parsec 
scale Broad Emission Line Region clouds - BELRs -  or Warm Absorbers 
- WAs) and distances from 1 pc (the size of the torus for an object with 
the non-beamed luminosity of Mkn~421, i.e. $L \sim 10^{42}$ erg s$^{-1}$) 
down to 0.5-1 light day (the size of the BELR and, possibly, the WA) the 
ionization parameter $U$ at the illuminated face of the cloud, during 100 
mCrab outbursts of the source, would reach value ranging between 
$U_{torus} \simeq 30$ and $U_{BELR-WA} \simeq 0.3-30$, putting most of 
the oxygen in the form of OVII and OVIII (for the BELR cloud) or OVIII 
and OIX (for the WA or the torus; see e.g. Figure 1 in Nicastro et al., 
1999). In the spectrum of Mkn~421 we only detect OVII, and measure 
upper limits for OVIII implying $U \ls 1$, so absorbing material from 
either the molecular torus or a WA in the nuclear environment of 
Mkn~421 along our line of sight is unlikely to produce the observed 
features. A particularly dense BELR cloud, instead, could. However, 
typical equivalent H column density for the BELR clouds are in the 
range $10^{22}-10^{24}$ cm$^{-2}$ (e.g. Blandford, et al., 1990), 
and metallicities are typically at least solar (e.g. Korista et al., 
1996). This would imply OVII columns of $> 2 \times 10^{18}$ cm$^{-2}$ 
(assuming an OVII fraction of $f_{OVII} = 0.2$ at $U_{BELR} = 0.3$, 
see Figure 1 of Nicastro et al., 1999), more than 3 orders of magnitude 
larger than the value we actually measure. 

The same argument can be used to estimate the ionization degree of 
an extremely high velocity ISM cloud, located at kpc-scale from the blazar's 
host galaxy center (and along the line of sight to Mkn~421). 
Assuming typical ISM densities of $\ls 1$ cm$^{-3}$ we obtain $U \gs 30$ 
during outburst phases, which implies virtually no OVII, with $f_{OIX} =
0.95$ and $f_{OVIII} = 0.05$ (e.g. Nicastro et al., 1999). 

Finally we also looked for variability of the strongest lines of these two 
absorption systems in the two separate ACIS-LETG and HRC-LETG spectra of 
Mkn~421, taken about 6 months apart and in different flux states. 
Variability of the absorption lines found here would denote an AGN origin. 
The OVII absorber at $z=0.011$ is clearly present, even if at lower 
significance, in both the ACIS-LETG and the HRC-LETG spectra and with same 
$W \sim 3$ m\AA, If the absorber was intrinsic to the nucleus, and if 
its density was $n_e \gs 10^6$ cm$^{-3}$ (Nicastro et al., 1999) the change 
in ionizing flux would have produced detectable opacity variation in the 
OVII K$\alpha$ line, which instead was not observed. 
For the $z=0.027$ system, the OVII lines looks stronger in the ACIS-LETG 
than in the HRC-LETG spectrum although their EWs of $\sim 2.5$ 
m\AA\ and $\sim 1.8$ m\AA\ respectively, are consistent with each others 
within the large $\pm 1$ m\AA\ 1$\sigma$ error. Unfortunately for the NVII 
and NVI lines, the strongest lines of this system, a comparison cannot be 
made, since only the HRC-LETG spectrum has enough counts to be sensitive to 
the detected columns: other HRC-LETG spectra with similar sensitivity 
limits are needed to strengthen this test.  

We conclude that the WHIM identification is the most likely for both the 
observed $z=0.011$ and $z=0.027$ systems. 

\subsection{Number Density and Mean Cosmological Mass Density of WHIM 
Filaments}
Despite the limited statistics of just two filaments detected along a 
single line of sight, our two detections of WHIM filaments along the line of 
sight to Mkn~421 allow us to obtain the first crude estimate of the number 
density of WHIM filaments per unit redshifts above a threshold column 
density, and mean cosmological mass density (Nicastro et al., 2005). 
The expected number density of WHIM filaments above a given ion column 
density is usually computed for the two most abundant ions of the oxygen, 
OVII and OVIII (e.g. Hellsten et al., 1998, Fang, Bryan, \& Canizares, 2002). 
The smallest OVII column density that we measure, is N$_{OVII} = (7 \pm 3) 
\times 10^{14}$ cm$^{-2}$, from the WHIM filament at $z = 0.027$. 
Using this as a threshold, gives a number of OVII WHIM absorbers per unit 
redshift and with N$_{OVII} \ge 7 \times 10^{14}$ cm$^{-2}$, of $\Delta 
\mathcal{N} / \Delta z = 2 / 0.03 \simeq 67^{+88}_{-43}$ (Gehrels, 1986). 
This estimate of $d \mathcal{N}_{OVII} /dz$ is slightly above but consistent 
with theoretical predictions, which predict $\sim 30$ OVII WHIM filaments 
down to the N$_{OVII} \ge 7 \times 10^{14}$ cm$^{-2}$ threshold and up to 
$z=0.03$ (Fang, Bryan \& Canizares, 2002: from their Fig. 13). 
Figure 18 shows the predicted curves $\mathcal{N}_{OVII}(z)$ versus 
$z$, for three different $N_{OVII}$ column density thresholds: $10^{16}$ 
(solid curve), $4 \times 10^{15}$ (dotted curve) and $7 \times 10^{14}$ 
cm$^{-2}$ (dashed curve; Fang, Bryan \& Canizares, 2002). 
Our data point, $\mathcal{N}_{OVII}(z=0.03;N_{OVII} \ge 7 \times 10^{14})$ 
is also shown with $\pm 1\sigma$ errors, and is consistent with the 
predicted curve $\mathcal{N}_{OVII}(z;N_{OVII} \ge 7 \times 10^{14})$: dashed 
curve in Fig. 18. 
The number density of OVII absorbers with N$_{OVII} \ge 7 \times 10^{14}$ 
cm$^{-2}$ is between 1.2 and 8.3 times that of the corresponding number 
density of OVI intergalactic absorbers with N$_{OVI} \gs 1.2 \times 10^{13}$ 
cm$^{-2}$ (i.e. $\sim 60$ times thinner than the OVII filaments), of 
$d\mathcal{N}/dz = 19 \pm 3$ (Danforth \& Shull, 2005). 
%
\begin{figure}
\hspace{-0.8in}
\epsfbox{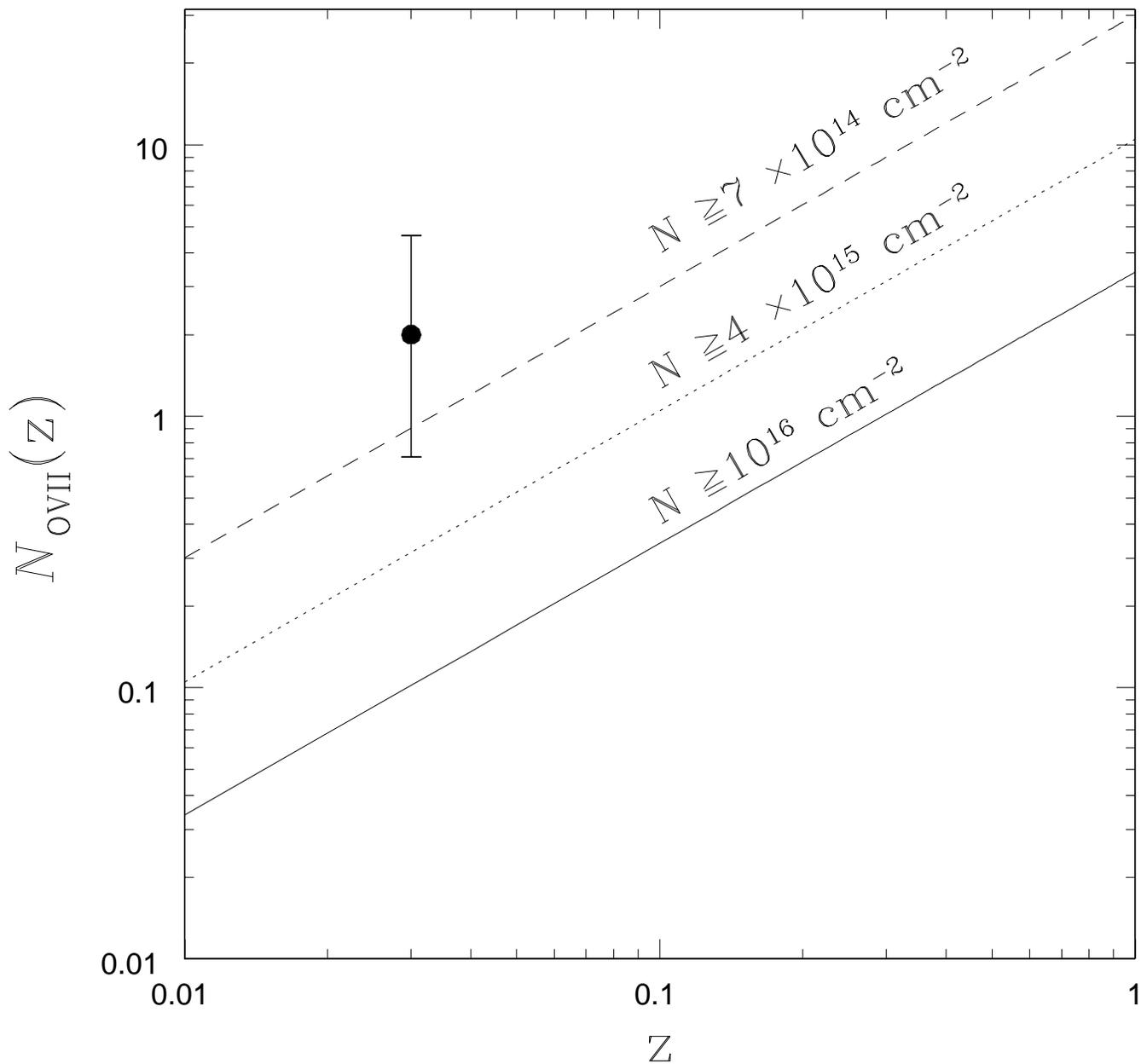}
\vspace{0in}\caption[h]{\footnotesize Predicted curves 
$\mathcal{N}_{OVII}(z)$ versus $z$, for three different $N_{OVII}$ column 
density thresholds: $10^{16}$ (solid curve), $4 \times 10^{15}$ (dotted 
curve) and $7 \times 10^{14}$ cm$^{-2}$ (dashed curve; from Fang, Bryan \& 
Canizares, 2002). 
The data point, $\mathcal{N}_{OVII}(z=0.03;N_{OVII} \ge 7 \times 10^{14})$ 
is also shown with $\pm 1\sigma$ errors, and is consistent with the 
dashed curve for $\mathcal{N}_{OVII}(z;N_{OVII} \ge 7 \times 10^{14})$.}
\end{figure}
%

Following Savage et al. (2002) (and approximating the formula to the 
low-redshift regime), we estimate the baryon mass density of the WHIM, in 
units of the current critical density $\rho_c$ as: 
$$\Omega_b = ({1 \over \rho_c}) ({{\mu m_p \sum_i  N_H^i} \over 
{d_{Mkn421}}}),$$ 
where $\mu$ is the average molecular weight (taken to be 1.3), $N_H^i$ is the 
equivalent H column density of the WHIM filament $i$ and $d_{Mkn421} 
\simeq (zc/H_0) = 128 h_{70}^{-1}$ Mpc, is the distance to Mkn~421. Averaging 
on the lower and upper limits of $N_H$ for the two filaments (\S 7.4, 7.5), 
and including low 
number statistics errors for a Poissonian distribution (Gehrels, 1986) gives 
$\Omega_b = 0.027^{+0.038}_{-0.019} \times 10^{-[O/H]_{-1}}$. Comparing 
this value with the $\Omega_b$ expected shows that virtually all the 
'missing' baryons at low redshift are to be found in the WHIM (Nicastro et 
al., 2005). 

\section{Summary and Conclusions}
In this paper we presented the first two high significance detections of 
X-ray WHIM filaments, in the very high signal to noise {\em Chandra}-LETG 
spectrum of the blazar Mkn~421 ($z = 0.03$). A recently published companion 
paper (Nicastro et al., 2005) focus on the measurement of the cosmological 
mass density in the OVII WHIM, derived from the detections presented here. 
Our main results are here summarized: 

\begin{itemize}
\item{} We detect weak metal absorption from two intergalactic ionized 
absorbers along the line of sight to Mkn~421, at average redshifts of 
$cz = 3300 \pm 300$ km s$^{-1}$ ($z=0.011$) and $cz = 8090 \pm 300$ km 
s$^{-1}$ ($z=0.027$). 

\item{} The $z=0.011$ X-ray absorber is detected through NVII, OVII and 
OVIII absorption, while the $z=0.027$ X-ray absorber is detected through 
CVI, NVI, NVII, OVII and NeIX. 

\item{}The conservative significances estimated for the $z=0.011$ and 
$z=0.027$ X-ray absorbers are 3.5$\sigma$ and 5.8$\sigma$. 

\item{} \underline{The $z=0.011$ system} is found at a redshift consistent 
with a known HI $Ly\alpha$ absorber, in a sub-dense galaxy environment 
about 5 times less populated 
than the average in the local Universe. 
The 3$\sigma$ upper limit temperature of the HI absorber, as derived from the 
observed 3$\sigma$ FWHM of the HI Ly$\alpha$, is inconsistent with the 
allowed range 
of temperatures of the X-ray absorber. If the two absorbers are related, 
they imply a multiphase IGM. 
The OVI detection in the FUSE spectrum of Mkn~421 at $z=0.011 \pm 0.001$ 
is hampered by a broad detector flaw. 

\item{} \underline{The $z=0.027$ system} lies $\sim 13$ Mpc from Mkn~421, 
and is found in an average density galaxy environment. 
No HI Ly$\alpha$ or OVI doublet are detected in the GHRS and FUSE 
spectra of Mkn~421 at $z=0.027$, down to the 3$\sigma$ sensitivity limits of 
N$_{HI} \ge 8.5 \times 10^{12}$ cm$^{-2}$ and N$_{OVI} \ge 1.4 \times 
10^{13}$ cm$^{-2}$. 

\item{} We find that a scenario in which the gas is primarily collisionally 
ionized, but still undergoing additional, and marginal, photoionization by 
the  diffuse UV-X-ray background and the proximity effect of Mkn~421 during 
bursts, allow us to define ranges of temperatures and metallicity 
ratios for both intervening systems. 

\item{} For the $z=0.011$ absorber we find a temperature range 
$T = (0.6-2.5) \times 10^6$ K. 
A rather tight lower limit can be set on the metallicity of this IGM system, 
of [O/H]$\ge -1.47$. 
Only a weak lower limit can be set on the relative N/O ratio, of [N/O]$ 
\ge -2.05$, consistent with the Solar value. 
Given the allowed range of temperature for the $z=0.011$ absorber, 
equivalent H column densities of N$_H = [(1.7 \pm 0.7)T_6 - (0.1 \pm 0.6)] 
\times 10^{19} \times 10^{-[O/H]_{-1}}$ (with [N/O] = 0) are implied. 
These in turn  imply a thickness of the absorber along our line of sight of 
$D = [(0.55 \pm 0.23)T_6 - (0.03 \pm 0.19)] \times 10^{-[O/H]_{-1}} 
(n_b)_{-5}^{-1}$ Mpc. 

\item{} For the $z=0.027$ absorber, we find that solution exist in the hybrid 
equilibrium collisional/photoionization scenario and for $n_b = 10^{-5}$ 
cm$^{-3}$, only for a narrow ($\Delta T \sim 2 \times 10^5$ K) 
range of temperature around T$=1.4 \times 10^6$ K, and only for rather high 
N/O ratios compared to Solar: [N/O]$\ge 0.65$, and moderately high 
metallicity [O/H]$\ge -1.32$. 
The breadth of the $\Delta T$ interval, around the T$=1.4 
\times 10^6$ K solution, depends on the 
metallicity ratio: values of [O/H] larger than -1 will broaden the 
allowed range of temperatures. 
For [N/O] = 0.65 we find N$_H = [(1.67 \pm 0.76)T_6 
+ (0.71 \mp 1.15)] \times 10^{19} \times 10^{-[O/H]_{-1}}$ cm$^{-2}$. 
This implies $D = [(0.53 \pm 0.24)T_6 + (0.23 \mp 0.37)] \times 
10^{-[O/H]_{-1}} (n_b)_{-5}^{-1}$ Mpc, assuming constant baryon 
density. 

\item{} All the physical properties of these two intervening absorbers 
are consistent with theoretical predictions for the WHIM. 
We therefore identify these two intervening absorbers with two tenuous WHIM 
filaments absorbing the X-ray radiation of Mkn~421 along our line of sight. 
This represents the first $> 3\sigma$ detection of WHIM filaments in the 
X-rays. 

\item{} Based on our detections we set for the first time an estimate of the 
number density of OVII WHIM filaments with N$_{OVII} \ge 7 \times 10^{14}$ 
cm$^{-2}$, in the local Universe, of $d\mathcal{N}/dz = 67^{+88}_{-43}$, 
(1.2-8.3) times that of the corresponding, but $\sim 60$ times 
thinner, OVI absorbers. Finally, from our equivalent hydrogen column density 
measurements, we derive a mean cosmological mass in X-ray WHIM filaments, 
in units of the current critical density of the Universe, of $\Omega_b = 
0.027^{+0.038}_{-0.019} \times 10^{-[O/H]_{-1}}$, consistent with both model 
predictions and the estimated number of 'missing' baryons at low redshift. 
\end{itemize}

This work only begins the study of the Warm-Hot IGM outside the Local group. 
The large number of ion species showing up in the X-ray spectra imply a 
rich field of investigation. Higher significance detections, along several 
other lines of sight, and at higher redshift, of He-like and H-like 
transitions from C, N, O and Ne, will allow us to measure the relative 
metallicity ratios at different redshifts. 
This will fundamentally contribute to the assessment of the still very poor 
measurements of metal production at relatively high redshift, and will help 
refining models for galaxy-IGM feedback, possibly providing invaluable 
hints for the main mechanisms responsible for this feedback (i.e. galaxy 
superwinds, versus quasars winds, etc.). Moreover, the potential for the 
simultaneous detection, in the same band, of inner-shell transition from 
lower ionized species of the same ions, will allow us to distinguish 
between different ionization scenarios, and to assess the importance and 
frequency of multiphase IGM. Reconstructing the distribution functions of 
several ions in the WHIM (i.e. $d\mathcal{N}/dzdN$), will also probe very 
useful in distinguishing different cosmologies (Fang \& Canizares, 2000), 
and possibly tracing the distribution of dark-matter potential wells in 
the local Universe. 

This work clearly shows that an extremely efficient way to detecting the 
WHIM with {\em Chandra} and {\em Newton}-XMM, is observing sources 
while in bright outburst phases. Mkn~421 is unique: no other extragalactic 
sources in the sky (except Gamma-Ray Burst X-ray afterglows, e.g. Fiore et 
al., 2000) reach the level of Mkn~421 during the observations presented here. 
However, as shown in Fig. 18, the expected number of 
OVII WHIM filament down to a given N$_{OVII}$ threshold, increases linearly 
with redshift. About 2 OVII filaments with N$_{OVII} \ge 10^{16}$ cm$^{-2}$ 
are expected up to $z=0.5$ along a random line of sight. Such OVII columns 
can easily be detected with 100 ks {\em Chandra} and/or {\em Newton}-XMM 
observations of background sources flaring at a level even 10-50 times lower 
than that reached by Mkn~421 during the observations presented here, 
or 100-500 ks observations of pre-selected $z \gs 0.3$ blazars with 
quiescent soft X-ray flux level of F$_{0.5-2 keV} \sim 1-2$ mCrab. 

A much greater potential is offered by the large collecting area of  
the spectrometers on the two planned X-ray 
missions, {\em Constellation}-X
\footnote{http://constellation.gsfc.nasa.gov/science/index.html}
and {\em XEUS}
\footnote{http://www.rssd.esa.int/index.php?project=XEUS}
. {\em Constellation}-X will 
allow us to detect WHIM systems as weak as those presented in this work, 
for possibly tens of different lines of sight, and thicker filaments 
will be detected towards hundreds of different sightlines. 
A resolving power of R$\sim 500$ at 20 \AA\
\footnote{http://constellation.gsfc.nasa.gov/science/design/spectral\_resolution.html}
will not allow us, in many cases, to resolve the lines and 
detect multiphase structures in the IGM. This will in turn hamper 
our capabilities of constraining the physical properties of the 
baryons in this diffuse web of WHIM and so of measuring $\Omega_b^{WHIM}$. 
A resolving power R$\gs 3000-5000$ is needed to resolve thermally broadened 
OVII lines in gas at $T=10^6$ K and so to (a) disentangle the Doppler 
parameter $b$ from the ion column density, and so derive the 
saturation-corrected contribution to $\Omega_b$ (b) measure the intrinsic 
internal turbulence in the filaments, and compare it to the galaxy velocity 
in the field, (c) allow for separation of multi-phase components, and in 
general, for a clear physical diagnostics (Elvis \& Fiore \& the {\em Pharos} 
Team, 2003). 

\begin{center}
{\bf Acknowledgments}
\end{center}
The authors thank the anonymous referee, for useful comments 
and suggestions that helped improve the paper. 
We thank Ron Remillard and the {\em Rossi}-XTE ASM team, for 
providing us with daily ASM flux notification for a list of bright 
blazars in the X-ray sky. 
We also thank the {\em Chandra} Mission Planning (MP) and Operations 
(Ops) teams for the rapid turn around and data processing. 
FN acknowledges support from the {\em Chandra} grant GO3-4152X and the 
LTSA grant NNG04GD49G.

\end{document}